\documentclass[10pt,a4paper]{article}
\usepackage[utf8]{inputenc}
\usepackage{a4wide}
\usepackage{graphicx}
\usepackage{hyperref}
\usepackage{color}

\usepackage{amssymb}
\usepackage{amsmath}
\usepackage{verbatim} 

\begin{document}

\begin{titlepage}
\begin{flushright}
\end{flushright}
\vfill
\begin{center}
{\Large\bf 	New Series Representations for the Two-Loop Massive Sunset Diagram}
\vfill
{\bf B. Ananthanarayan$^a$, Samuel Friot$^{b,c}$ and Shayan Ghosh$^d$}\\[1cm]
{$^a$ Centre for High Energy Physics, Indian Institute of Science, \\
Bangalore-560012, Karnataka, India}\\[0.5cm]
{$^b$ Institut de Physique Nucl\'eaire d’Orsay, Universit\'e Paris-Sud 11, \\
IN2P3-CNRS, F-91405 Orsay Cedex, France} \\[0.5cm]
{$^c$ Institut de Physique des 2 Infinis de Lyon, Universit\'e Lyon 1, \\
IN2P3-CNRS, F-69622 Villeurbanne Cedex, France}\\[0.5cm]
{$^d$ Helmholtz-Institut für Strahlen- und Kernphysik, Universität Bonn,}\\
53115 Bonn, Germany \\
\end{center}
\vfill
\begin{abstract}
We derive new convergent series representations for the two-loop sunset diagram with three different propagator masses $m_1,\, m_2$ and $m_3$ and external momentum $p$ by techniques of
analytic continuation on a well-known triple series that corresponds to the Lauricella $F_C^{(3)}$ function. The convergence regions of the new series contain regions of interest to physical problems. These
include some ranges of masses and squared external momentum values which make them useful from Chiral Perturbation Theory to some
regions of the parameter space of the Minimal Supersymmetric Standard Model. The analytic continuation results presented on the Lauricella series could be used in other settings as well.

\end{abstract}
\vfill
\vfill
\end{titlepage}

\section{Introduction}
\begin{figure}[hbtp]
\centering
\includegraphics[scale=0.7]{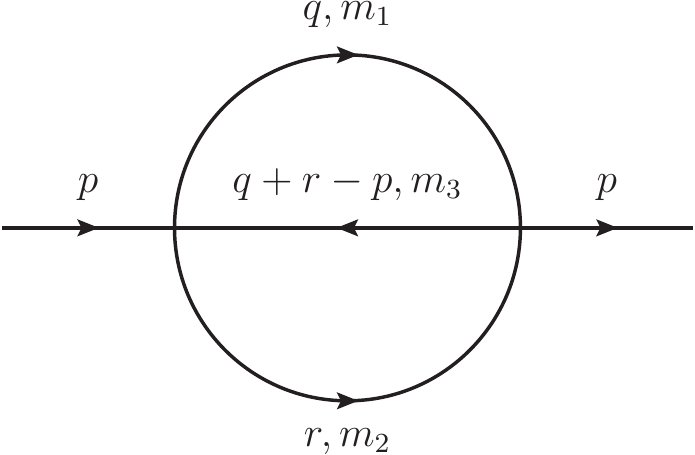}
\caption{The sunset diagram}
\label{FigSunset}
\end{figure}

The sunset integral is among the simplest of two-loop integrals that appear in the perturbation expansion of quantities in various quantum field theories, including
the Standard Model (SM).
In the convention of the classic work \cite{Berends:1993ee}, the general massive sunset diagram of Fig.~\ref{FigSunset} is defined as
\begin{align}
	T(p^2, m_1^2, m_2^2, m_3^2,\alpha, \beta, \gamma) \equiv \frac{1}{i^2\pi^4}\int \frac{d^nq}{(2\pi\mu)^{n-4}} \frac{d^nr}{(2\pi\mu)^{n-4}} \frac{1}{\left[ q^2 - m_1^2 \right]^{\alpha} \left[ r^2 - m_2^2\right]^{\beta} \left[  (q+r-p)^2 - m_3^2 \right]^{\gamma}} 
\end{align}
where the external momentum $p^2=s$ can take any value, $n=4-2\epsilon$
and $m_1,\, m_2,\, m_3$ are the masses of the three distinct particles in the propagators. 
The most general sunset integral, therefore, can have up to four independent mass scales. The sunset integral with general powers of the propagators $(\alpha,\beta,\gamma)$ can be reduced using integration by parts into a linear combination of a maximum of four master integrals (MI) \cite{Tarasov:1997kx}. These master integrals are the sunset integral with the following propagator powers: $(\alpha,\beta,\gamma) = (1,1,1),(2,1,1),(1,2,1),(1,1,2)$. 
It may also be noted that all tensor and vector integrals, defined with momentum factors and
appropriate Lorentz indices inside the integrals, as well as arbitrary derivatives with respect
to the momentum and / or the masses can also be reduced to the same set of MI.  Thus
the study of the MI becomes of great importance in the context of the sunset integrals.
Important special cases for renormalization theory include the fact
that at $p^2=0$ the residues in dimensional regularization can be evaluated in closed form,
see \cite{DT,CCLR}.  For configurations known as threshold and pseudo-threshold configurations
the coefficient of the $\epsilon^0$ is also known in terms of elementary functions, see
\cite{BDU}. Despite these early important advances,
much work continues to be done on the evaluation of the sunset integrals.  
In particular, concerning the analytical evaluation of the sunset and more generally of the two-loop self energies, an implementation of analytical results is still 
far from complete, and their evaluation remains often numerical, 
 see for example the recent phenomenological work of \cite{Borowka:2018anu} in the context
of the Minimal Supersymmetric Standard Model (MSSM). 

As far as the analytic evaluation of the sunset diagrams is concerned
we will not try to give a list of 
more recent publications dedicated to the sunset here but instead 
refer the reader to the extensive bibliography in \cite{Adams:2015gva} 
(see also the list given in \cite{Borowka:2018anu}). In \cite{Adams:2015gva} the most general 4 mass scale sunset integral has been evaluated in terms of newly defined generalized elliptic functions. In the classic work \cite{Berends:1993ee}, the most general 4 mass scale sunset integral has also been calculated analytically in terms of Lauricella multiple series.  These can be obtained from introducing series
representations for the Bessel functions in terms of which the propagators can be expressed, or by
using among others Mellin-Barnes (MB) representations.  It may be noted here that 
the analytic expressions given 
in \cite{Berends:1993ee} can be used only for restricted ranges of the 
values of the masses and squared external momentum values ($p^2$) 
derived from the convergence regions of the Lauricella series.  In what follows,
we will refer to these as known series representations of the general four mass scale sunset.

It may also be noted here that the sunset integrals also appear in effective field theories
of the SM.  One of the important effective field theories of the SM at low-energies is
Chiral Perturbation Theory (ChPT), the effective theory of the pseudo-scalar octet of the
light mesons associated with the spontaneous symmetry breaking of the approximate axial-vector
symmetry, part of the symmetry associated with independent left- and right- chiral rotations
of the QCD Lagrangian when the three lightest quark masses are sent to 0.  The pseudo-scalar
octet in the absence of iso-spin breaking and electromagnetic corrections has three masses,
namely the pion, kaon and eta masses.  At two-loop order, the masses and the decay constants
of these particles involve sunset diagrams that themselves require the masses to be inserted
into the two-loop integrals, see \cite{ABT}.

Driven by the needs of ChPT, in recent works \cite{Ananthanarayan:2017qmx, Ananthanarayan:2018irl} (see also \cite{Ananthanarayan:2016pos,Ananthanarayan:2017yhz}), analytic representations of the masses and decay constants of the light pseudoscalar mesons have been derived at the two-loop level. The relevant sunset diagrams that appear here have special values of the masses $m_1, m_2$ and $m_3$ and squared external momentum $p^2$
(all in terms of $m_\pi, m_K$ and $m_\eta$) that do not belong to the convergence regions of the analytic expressions given in \cite{Berends:1993ee}. One of the aims of the works 
\cite{Ananthanarayan:2017qmx, Ananthanarayan:2018irl} was to give the ChPT results in terms of series representations which are easy to handle and computationally efficient at the same time. Therefore, to get the required results in \cite{Ananthanarayan:2017qmx, Ananthanarayan:2018irl} we have followed the spirit of \cite{Berends:1993ee}. However, instead of considering the most general two-loop sunset integral as in \cite{Berends:1993ee}, we have focused on the particular sunset configurations needed in ChPT: we have developed the Mellin-Barnes approach to evaluate the ChPT sunsets and analytical results were presented in terms of double series of Kamp\'e de F\'eriet type (details of this derivation will be given in \cite{AFGkaon}).

In the present work we return to the most general massive sunset and we show that, as discussed in the conclusion and outlook section of \cite{Berends:1993ee}, an analytic continuation procedure applied to the expressions presented in terms of Lauricella series in the latter paper allows one to provide new series representations of the sunset diagram, giving access to some ranges of values of the masses and squared external momentum which are not reachable using the results of \cite{Berends:1993ee}. Let us mention that these new ranges of values include, among others, the masses and squared external momentum needed in the ChPT context, which gives a cross-check of the results given in \cite{Ananthanarayan:2017qmx}, \cite{ Ananthanarayan:2018irl} and \cite{AFGkaon}. The analytic continuation expressions that will be presented here are however more general than the ChPT results since $p^2$ is a free parameter whereas in ChPT it is fixed to the value of one of the squares of the meson masses, as we have discussed above.
Therefore, we will see that our formulas may be applied in completely different contexts, and an illustration will be presented in the MSSM case \cite{Borowka:2018anu}. Of interest also in physical applications is the expansion of the formulas as Laurent series in $\epsilon$. We will discuss this point too, at any order in $\epsilon$. However, to avoid very lenghty expressions, this will be detailed on a single particular example only.

This paper is organized as follows: in Section \ref{BERENDSetAl}, we recall the well-known representations of the 4-mass scale sunset diagram in terms of Lauricella functions. In Section \ref{ANALYTIC_C}, we develop the analytic continuation procedure on the latter, to obtain new series representations of the sunset diagram that are presented in Section \ref{new} and in the Appendix. The $\epsilon$-expansion procedure is briefly shown in Section \ref{eps}. Then, in Section \ref{physics}, some physical situations in ChPT and in the MSSM are considered and we show which series can be applied to them. Some conclusions follow.

\section{Some known series representations of the sunset diagram\label{BERENDSetAl}}
Let us begin by recalling some known series representations of the
sunset integral, first presented in \cite{Berends:1993ee} where small and large momentum expansions have been derived for the sunset MI $T(p^2, m_1^2, m_2^2, m_3^2, 1, 1, 1)$.

The small momentum expansion of the sunset diagram $T(p^2, m_1^2, m_2^2, m_3^2, 1, 1, 1)$ is given as \cite{Berends:1993ee} 
\begin{align}
&S_1\equiv-m_3^2\left(\frac{m_3^2}{4\pi\mu^2}\right)^{2(\nu-1)}\nonumber \\ 
&\times\left\{
z_1^\nu z_2^\nu \Gamma^2(-\nu)F_C^{(3)}(1, 1+\nu; 1+\nu, 1+\nu, 1+\nu;z_1, z_2, z_3)\right. \nonumber \\
&-z_1^\nu \Gamma^2(-\nu)F_C^{(3)}(1, 1-\nu; 1+\nu,1-\nu, 1+\nu; z_1, z_2, z_3) \nonumber \\  
&-z_2^\nu \Gamma^2(-\nu)F_C^{(3)}(1, 1-\nu; 1-\nu, 1+\nu, 1+\nu; z_1, z_2, z_3) \nonumber 
\end{align}
\begin{align}
&\left.-\Gamma(\nu) \Gamma(-\nu)\Gamma(1-2\nu)F_C^{(3)}(1-2\nu, 1-\nu;1-\nu, 1-\nu, 1+\nu; z_1, z_2, z_3) \right\}\label{Berends}
\end{align}
where 
\begin{align}\label{zi}
z_1=\frac{m_1^2}{m_3^2},\ z_2=\frac{m_2^2}{m_3^2},\ z_3=\frac{p^2}{m_3^2}\ \ \textrm{and}\ \ \nu=1-\epsilon\ .
\end{align}

\begin{figure}[hbtp]
\centering
\includegraphics[scale=0.4]{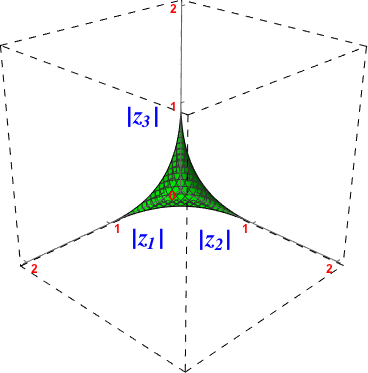}
\caption{Convergence region of the Lauricella triple series in Eq.(\ref{Lauricella}).}.
\label{ConvLauricella}
\end{figure}

In Eq.(\ref{Berends}) appears the well-known Lauricella triple series
\begin{align}
	F_C^{(3)}(a,b;c,d,e;z_1,z_2,z_3) = \sum_{m,n,p=0}^\infty \frac{(a)_{m+n+p} (b)_{m+n+p}}{(c)_m (d)_n (e)_p} \frac{z_1^m}{m!} \frac{z_2^n}{n!} \frac{z_3^p}{p!}\ ,
	\label{Lauricella}
\end{align}
where the superscript $^{(3)}$ indicates the number of its independent variables and $(\alpha)_m\equiv\Gamma(\alpha+m)/\Gamma(\alpha)$ is the Pochhammer symbol. In the most general case, the Lauricella series are generalizations of hypergeometric series to multiple variables and can be of four different types: $F_A^{(N)}, F_B^{(N)}, F_C^{(N)}$ and $F_D^{(N)}$ (see for instance \cite{Kampe, Exton1, Srivastava} for details) but it is the case of three variables that is commonly referred to as a Lauricella series. The convergence region of the triple series in Eq.(\ref{Lauricella}) is given by the inequality 
\begin{align}
\sqrt{|z_1|}+\sqrt{|z_2|}+\sqrt{|z_3|}<1
\label{conv1}
\end{align}
which is shown graphically in Fig.~\ref{ConvLauricella}. 

In this figure, the convergence region is shown in $\mathbb{R}_0^3$, where 
\begin{align}
\mathbb{R}_0\doteq\mathbb{R}_+\cup\{0\}\ ,
\end{align}
namely the first (closed) octant $(+++)$ of the $(z_1,z_2,z_3)$-space.
Indeed, since the convergence of an $N$-dimensional power series in $z_1,...,z_N$ is in general absolute, its convergence region $\mathcal{D}$, which is a subset of $\mathbb{C}^N$,
may be adequately described by the domain $\mathbf{D}$ in absolute space $\mathbb{R}_0^N$, obtained by the natural projection $(z_1,...,z_N)\rightarrow(\vert z_1\vert,...,\vert z_N\vert)$\cite{Srivastava}. The full convergence region (in the other octants) can be simply obtained by symmetry. 

In the rest of this paper, however, giving the convergence regions only in the first octant will not be sufficient, because the analytic continuation formulas will be expressed as power series not in the $z_i$ but in some intricate combinations of the latter. It will therefore not be obvious to deduce regions of convergence in the full space from the first octant only. However, in order to simplify the figures as much as possible, we will show in this paper only the first octant and the fifth (the latter being defined as $(++-)$), since these are the most relevant from the physics point of view (indeed, from Eq.(\ref{zi}) we see that $z_1$ and $z_2$ are positive quantities while $z_3$ may be either positive or negative).

\begin{figure}[hbtp]
\centering
\includegraphics[scale=0.4]{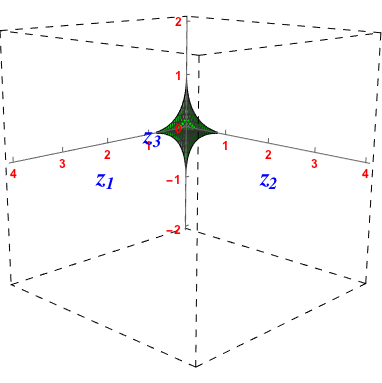}
\caption{Convergence region of the r.h.s. of Eqs.(\ref{Berends}) and (\ref{Lauricella}) in the first and fifth octants.}.
\label{ConvBerends}
\end{figure}

From Eqs.(\ref{zi}) and (\ref{conv1}), it is clear that each of the series of Eq.(\ref{Berends}) converges in the region 
\begin{align}
m_1+m_2+\sqrt{\vert p^2\vert}<m_3\ ,
\label{Conv_L1}
\end{align}
shown in Fig. \ref{ConvBerends}, which is the same as Fig. \ref{ConvLauricella} with the addition of the fifth octant ($z_1\geq 0$, $z_2\geq 0$, $z_3\leq 0$).

To derive the sunset expression for large-$p^2$ values, the authors of \cite{Berends:1993ee} have used the standard analytic continuation formula for the Lauricella $F_C^{(3)}$ series , namely \cite{Exton1}
\begin{align}
	F_C^{(3)}&(a,b;c,d,e;z_1,z_2,z_3)  \nonumber\\
	&=\frac{\Gamma(e)\Gamma(b-a)}{\Gamma(b)\Gamma(e-a)}(-z_3)^{-a}F_C^{(3)}\left(a,1+a-e;c,d,1-b+a;\frac{z_1}{z_3},\frac{z_2}{z_3},\frac{1}{z_3}\right)\nonumber\\
	&+\frac{\Gamma(e)\Gamma(a-b)}{\Gamma(a)\Gamma(e-b)}(-z_3)^{-b}F_C^{(3)}\left(b,1+b-e;c,d,1-a+b;\frac{z_1}{z_3},\frac{z_2}{z_3},\frac{1}{z_3}\right).
	\label{LauricellaAC}
\end{align}
The r.h.s of Eq.(\ref{LauricellaAC}) converges in the region
\begin{align}
\sqrt{|z_1|}+\sqrt{|z_2|}+1<\sqrt{|z_3|}\ .
\label{conv2}
\end{align}

\begin{figure}[hbtp]
\centering
\includegraphics[scale=0.4]{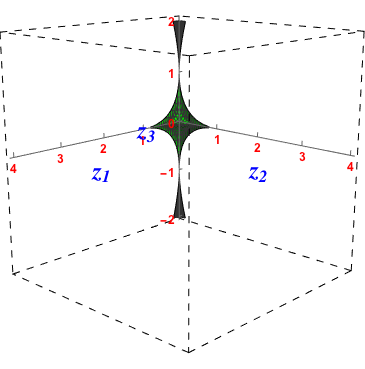}
\caption{Convergence regions of the r.h.s. of Eq.(\ref{BerendsAC}) (top and bottom $\vert z_3\vert>1$ regions) and of Eq.(\ref{Berends}) ($\vert z_3\vert<1$ regions).}.
\label{ConvBerends3}
\end{figure}

When Eq.(\ref{LauricellaAC}) is applied to the sunset representation given in Eq.(\ref{Berends}), one gets the following large momentum expansion of $T(p^2, m_1^2, m_2^2, m_3^2, 1, 1, 1)$ 
\begin{align}
&S_2\equiv -m_3^2\left(\frac{m_3^2}{4\pi\mu^2}\right)^{2(\nu-1)}\nonumber\\ 
\times &\left\{-\frac{z_1^\nu}{z_3} \Gamma^2(-\nu)F_C^{(3)}\left(1,1-\nu;1+\nu,1-\nu,1+\nu;\frac{z_1}{z_3},\frac{z_2}{z_3},\frac{1}{z_3}\right)\right. \nonumber\\
&-\frac{z_2^\nu}{z_3} \Gamma^2(-\nu)F_C^{(3)}\left(1,1-\nu;1-\nu,1+\nu,1+\nu;\frac{z_1}{z_3},\frac{z_2}{z_3},\frac{1}{z_3}\right)\nonumber\\
&-\frac{(z_1 z_2)^\nu}{z_3} \Gamma^2(-\nu)F_C^{(3)}\left(1,1-\nu;1+\nu,1+\nu,1-\nu;\frac{z_1}{z_3},\frac{z_2}{z_3},\frac{1}{z_3}\right)\nonumber\\
&+\frac{z_1^\nu}{z_3}(-z_3)^\nu \frac{\Gamma^2(-\nu)\Gamma(\nu)\Gamma(1+\nu)}{\Gamma(2\nu)}F_C^{(3)}\left(1-\nu,1-2\nu;1+\nu,1-\nu,1-\nu;\frac{z_1}{z_3},\frac{z_2}{z_3},\frac{1}{z_3}\right)\nonumber\\
&+\frac{z_2^\nu}{z_3}(-z_3)^\nu \frac{\Gamma^2(-\nu)\Gamma(\nu)\Gamma(1+\nu)}{\Gamma(2\nu)}F_C^{(3)}\left(1-\nu,1-2\nu;1-\nu,1+\nu,1-\nu;\frac{z_1}{z_3},\frac{z_2}{z_3},\frac{1}{z_3}\right)\nonumber
\end{align}
\begin{align}
&-(-z_3)^{\nu-1}\frac{\Gamma^2(-\nu)\Gamma(\nu)\Gamma(1+\nu)}{\Gamma(2\nu)}F_C^{(3)}\left(1-\nu,1-2\nu;1-\nu,1-\nu,1+\nu;\frac{z_1}{z_3},\frac{z_2}{z_3},\frac{1}{z_3}\right)\nonumber\\
&+\left.(-z_3)^{2\nu-1}\frac{\Gamma^3(\nu)\Gamma(1-2\nu)}{\Gamma(3\nu)}F_C^{(3)}\left(1-3\nu,1-2\nu;1-\nu,1-\nu,1-\nu;\frac{z_1}{z_3},\frac{z_2}{z_3},\frac{1}{z_3}\right)\right\}
\label{BerendsAC}
\end{align}
which is valid, due to Eq.(\ref{conv2}), for $\vert p^2\vert>(m_1+m_2+m_3)^2$. This region is shown in Fig. \ref{ConvBerends3}. 
Note that Eq.(\ref{BerendsAC}) slightly differs from Eq.(24) of \cite{Berends:1993ee}. Indeed, in the latter, terms like $(-z_3)^{-a}$ coming from Eq.(\ref{LauricellaAC}) were rewritten as $(-1/z_3)^{a}$, which is not true when $a$ is not an integer and $z_3>0$. Therefore we have preferred to leave these terms as they naturally appear when one applies Eq.(\ref{LauricellaAC}) on Eq.(\ref{Berends}). We have also corrected several misprints in the last two terms of Eq.(24) of \cite{Berends:1993ee}.

Eqs.(\ref{Berends}) and (\ref{BerendsAC}) are the main results of \cite{Berends:1993ee} for the sunset diagram (with unit powers of the propagators).

\subsection{Invariance of the sunset diagram under permutation of masses\label{invariance}}

Due to the invariance of the Lauricella series under the permutation of its variables (and of the corresponding parameters) one can obtain from Eq.(\ref{LauricellaAC}) the following alternative analytic continuation formula
\begin{align}
	F_C^{(3)}&(a,b;c,d,e;z_1,z_2,z_3)  \nonumber\\
	&=\frac{\Gamma(c)\Gamma(b-a)}{\Gamma(b)\Gamma(c-a)}(-z_1)^{-a}F_C^{(3)}\left(a,1+a-c;1-b+a,d,e;\frac{1}{z_1},\frac{z_2}{z_1},\frac{z_3}{z_1}\right)\nonumber\\
	&+\frac{\Gamma(c)\Gamma(a-b)}{\Gamma(a)\Gamma(c-b)}(-z_1)^{-b}F_C^{(3)}\left(b,1+b-c;1-a+b,d,e;\frac{1}{z_1},\frac{z_2}{z_1},\frac{z_3}{z_1}\right)
	\label{LauricellaAC2}
\end{align}
and an analogous relation derived from Eq.(\ref{LauricellaAC2}) by exchanging $(z_1,c)$ with $(z_2,d)$.

\begin{figure}[hbtp]
\centering
\includegraphics[scale=0.4]{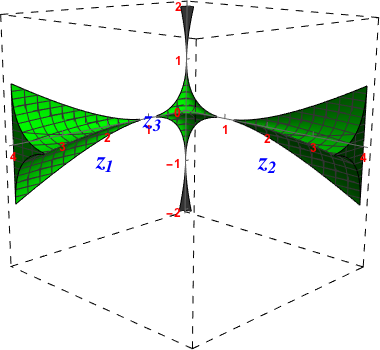}
\caption{Convergence regions given in Eq.(\ref{Conv2_L1}) (left region with $z_1>1$) and Eq.(\ref{Conv3_L1}) (right region with $z_2>1$), with the regions already shown in Fig.\ref{ConvBerends3} (central region).}
\label{Reg2}
\end{figure}

When these relations are used to replace each of the Lauricella series in Eq.(\ref{Berends}), they yield, after simplification,

\begin{align}
&S_3\equiv-m_1^2\left(\frac{m_1^2}{4\pi\mu^2}\right)^{2(\nu-1)}\nonumber \\ 
&\times\left\{
z_1^{-\nu} \left(\frac{z_2}{z_1}\right)^\nu \Gamma^2(-\nu)F_C^{(3)}\left(1, 1+\nu; 1+\nu, 1+\nu, 1+\nu;\frac{1}{z_1}, \frac{z_2}{z_1},\frac{z_3}{z_1}\right)\right.\nonumber \nonumber \\
&-z_1^{-\nu} \Gamma^2(-\nu)F_C^{(3)}\left(1, 1-\nu; 1+\nu,1-\nu, 1+\nu; \frac{1}{z_1}, \frac{z_2}{z_1}, \frac{z_3}{z_1}\right) \nonumber 
\end{align}
\begin{align} 
&-\left(\frac{z_2}{z_1}\right)^\nu \Gamma^2(-\nu)F_C^{(3)}\left(1, 1-\nu; 1-\nu, 1+\nu, 1+\nu; \frac{1}{z_1}, \frac{z_2}{z_1}, \frac{z_3}{z_1}\right) \nonumber \\
&\left.-\Gamma(\nu) \Gamma(-\nu)\Gamma(1-2\nu)F_C^{(3)}\left(1-2\nu, 1-\nu;1-\nu, 1-\nu, 1+\nu; \frac{1}{z_1}, \frac{z_2}{z_1}, \frac{z_3}{z_1}\right) \right\}\label{Berends2}
\end{align}
which converges for
\begin{align}
m_2+m_3+\sqrt{\vert p^2\vert}<m_1\ ,
\label{Conv2_L1}
\end{align}
and the following expression, similar to $S_3$
\begin{align}
S_4=S_3(m_1\leftrightarrow m_2, z_1\leftrightarrow z_2)\ ,
\label{S_4}
\end{align}
the convergence region of the latter analytic continuation is
\begin{align}
m_1+m_3+\sqrt{\vert p^2\vert}<m_2\ .
\label{Conv3_L1}
\end{align}
It is clear, however, that $S_3$ and $S_4$ are just $S_1$, given in Eq.(\ref{Berends}), if one uses the invariance property of the sunset diagram under the permutations of the masses
\[
  p_1 = 
  \begin{pmatrix}
    m_1 & m_2 & m_3 \\
    m_3 & m_2 & m_1 
  \end{pmatrix}
\]
and
\[
  p_2 = 
  \begin{pmatrix}
    m_1 & m_2 & m_3 \\
    m_1 & m_3 & m_2 
  \end{pmatrix}.
\]
Therefore, from the physics point of view, one cannot learn anything new from $S_3$ and $S_4$ compared to $S_1$, even if in the $(z_1,z_2,z_3)$-space the convergence regions of the former are, obviously, not the same as the one of the latter. For illustration, these regions are depicted in Fig. \ref{Reg2}, respectively to the left and to the right of the central convergence region that we have already shown in Fig. \ref{ConvBerends3}.

It should be noted that, contrary to the cases of $S_3$ and $S_4$ considered in the present section, the invariance of the Lauricella series under the permutation of its variables will give birth to non trivial series representations of the sunset in Section \ref{new}.

\begin{figure}[hbtp]
\centering
\includegraphics[scale=0.4,angle=1,origin=C]{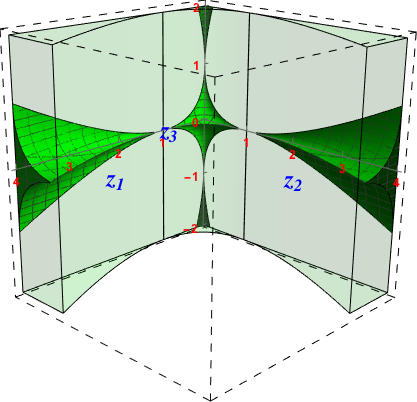}
\caption{Extended convergence region where a careful numerical treatment is needed (light green).}.
\label{ConvBerends4}
\end{figure}

 It is clear from Fig.~\ref{Reg2} that a large part of the $(z_1,z_2,z_3)$-space cannot be reached by any of the series representations that we have discussed so far. 
It is argued in \cite{Berends:1993ee} (although not quantitatively shown) that by analyticity the convergence region of the total sum in the r.h.s. of Eq.(\ref{Berends}) may be extended beyond the convergence region of each of its individual series, at the price of a careful numerical treatment at those points which belong to the former and not to the latter. This extended region is given by the values of the masses and external momentum that simultaneously satisfy the $\vert p^2\vert<(m_1+m_2+m_3)^2$ threshold condition and the condition $m_1+m_2<m_3$ (as well as those values for which the latter conditions are changed by permutations of the masses). It is shown in light green in Fig. \ref{ConvBerends4}.

The aim of the present work is to find new analytic continuation formulas for the Lauricella $F_C^{(3)}$ series in order to derive from them some series representations of the sunset diagram valid in the domain of the $(z_1,z_2,z_3)$-space where none of the series representations presented until here can be used. We also wish to access the light green region of Fig. \ref{ConvBerends4} without the careful numerical treatment mentioned in \cite{Berends:1993ee}, and with a robust analytic method.

\section{New analytic continuations for the Lauricella $F_C^{(3)}$ series\label{ANALYTIC_C}}

To derive analytic continuation formulas for a multiple series, one possibility is to write some of the infinite sums of the multiple series in terms of a relatively common function for which analytic continuations are known or can be obtained, replace this function by one of these analytic continuations, and find the convergence region of the newly obtained multiple series, hoping that it will extend beyond the convergence region of the initial multiple series. This is the strategy that we adopt here for the analytic continuation of the Lauricella series and, more precisely, we will rewrite the latter as an infinite sum of Appell $F_4$ double hypergeometric functions. We indeed have derived new analytic continuation formulas for the Appell $F_4$ function in \cite{AFGH}. Note however that instead of making the Appell $F_4$ function appear in our calculations, one could have chosen to write single sums as ${}_pF_q$ hypergeometric functions and performed a similar analysis (see \cite{PhD_Shayan} for detailed calculations with this alternative approach). In general, the analytic continuation of a multiple series is a sum of several multiple series and the convergence region of the resulting expression is given by the intersection of the convergence regions associated with each of the multiple series obtained at the end. It may also be necessary, for the validity of the new expression, to take into account the constraints coming from the range of validity of the analytic continuation formulas or other relations that have been used in the intermediate steps of the derivation (if any).

As a first simple illustration of this general procedure, we will now derive Eq.(\ref{LauricellaAC}) from Eq.(\ref{Lauricella}). To this end, one can either write one of the three sums of the Lauricella $F_C^{(3)}$ series as a Gauss ${}_2F_1$ hypergeometric series, and then use a well-known analytic continuation formula of ${}_2F_1$ or, equivalently, one can write two sums of the Lauricella series as an Appell $F_4$ series and use a known analytic continuation formula of the latter. 
Let us briefly demonstrate the second way. 

Let us write
\begin{align}
F_C^{(3)}(a, b; c, d, e;z_1, z_2, z_3)&=\sum_{m=0}^\infty\frac{(a)_m(b)_m}{(c)_m}\frac{z_1^m}{m!}\ F_4(a+m,b+m;d,e;z_2,z_3)
	\label{FC3_F4}
\end{align}
where 
\begin{align}
F_4(\alpha,\beta;\gamma,\delta;x,y)=\sum_{i,j=0}^\infty\frac{(\alpha)_{i+j}(\beta)_{i+j}}{(\gamma)_i(\delta)_j}\frac{x^i}{i!}\frac{y^j}{j!}.
	\label{F4}
\end{align}
We now replace the Appell $F_4$ series in Eq.(\ref{FC3_F4}) using the analytic continuation formula \cite{Kampe}
\begin{align}
	F_4(\alpha,\beta;&\gamma,\delta;x,y) = \nonumber \\
	&\frac{\Gamma(\delta)\Gamma(\beta-\alpha)}{\Gamma(\delta-\alpha)\Gamma(\beta)} (-y)^{-\alpha}
	 F_4\left(\alpha,\alpha-\delta+1;\gamma,\alpha-\beta+1;\frac{x}{y},\frac{1}{y}\right) \nonumber \\
	 +&\frac{\Gamma(\delta)\Gamma(\alpha-\beta)}{\Gamma(\delta-\beta)\Gamma(\alpha)} (-y)^{-\beta}
	 F_4\left(\beta,\beta-\delta+1;\gamma,\beta-\alpha+1;\frac{x}{y},\frac{1}{y}\right)\ .
	 \label{F4_AC1}
\end{align}
Note that $F_4=F_C^{(2)}$, therefore Eq.(\ref{F4_AC1}) is in fact the double series analogue of Eq.(\ref{LauricellaAC}).

Rewriting both $F_4$ functions of the r.h.s. in terms of their series representations following Eq.(\ref{F4}), and using the relation
\begin{align}
	\Gamma(z-n)=\frac{\Gamma(z)\Gamma(1-z)(-1)^n}{\Gamma(n+1-z)}\ ,\ \ \text{for}\ \  n\in \mathbb{Z}\ ,
	 \label{reflection}
\end{align}
one may then recognize Lauricella $F_C^{(3)}$ series in each of the two terms which appear at the end. This yields Eq.(\ref{LauricellaAC}).

As a remark, let us mention that one can prove Eq.(\ref{F4_AC1}) in a similar way to what we just did, using the well-known analytic continuation of the Gauss hypergeometric function
\begin{align}
	{}_2F_1(\alpha,\beta;\gamma;x) =
	&\frac{\Gamma(\gamma)\Gamma(\beta-\alpha)}{\Gamma(\gamma-\alpha)\Gamma(\beta)} (-x)^{-\alpha}
	 {}_2F_1\left(\alpha,\alpha-\gamma+1;\alpha-\beta+1;\frac{1}{x}\right) \nonumber \\
	 +&\frac{\Gamma(\gamma)\Gamma(\alpha-\beta)}{\Gamma(\gamma-\beta)\Gamma(\alpha)} (-x)^{-\beta}
	 {}_2F_1\left(\beta,\beta-\gamma+1;\beta-\alpha+1;\frac{1}{x}\right)
	 \label{2F1_AC}
\end{align}
and, as said above, it is also possible to directly derive Eq.(\ref{LauricellaAC}) by inserting Eq.(\ref{2F1_AC}) in the Lauricella series written as a double sum of Gauss hypergeometric series.

It is remarkable that the r.h.s. of the analytic continuation formula Eq.(\ref{LauricellaAC}) (Eq.(\ref{F4_AC1})) is expressed only in terms of the same type of Lauricella function (Appell function) as its l.h.s\footnote{Notice however that this nice property does not hold for non-trivial analytic continuations and, in fact, not even when one applies to above simple analysis on the other Lauricella series $F_A^{(3)}$, $F_B^{(3)}$ and $F_D^{(3)}$ (analogously for the corresponding Appell series).}. This gives the convergence region of the r.h.s in a straightforward way, since it is sufficient to replace $(z_1,z_2, z_3)$ in Eq.(\ref{conv1}) by $(\frac{z_1}{z_3},\frac{z_2}{z_3},\frac{1}{z_3})$. Doing this, we obtain Eq.(\ref{conv2}). 

Concerning Feynman integrals and other physical quantities of interest that are expressed as a sum of several different multiple series, one has to apply the analytic continuation procedure to each of the involved multiple series that need it, and the complexity of the intermediate steps can differ significantly from one multiple series to the other. However, in the case of the sunset, we have seen that the results given in \cite{Berends:1993ee} are written in terms of only one type of multiple series, the Lauricella $F_C^{(3)}$, therefore one does not face this difficulty. Indeed, once relevant analytic continuations for the $F_C^{(3)}$ Lauricella series are obtained, it will be possible to apply them to each of the different terms of Eqs.(\ref{Berends}), (\ref{BerendsAC}), (\ref{Berends2}) and to the symmetrical equations obtained by the exchange of the variables (and corresponding coefficients) of the latter. Doing so, we will derive several new series representations of the sunset, which will greatly enlarge the reachable region of the $(z_1,z_2,z_3)$-space shown in Fig. \ref{Reg2}. In this obtained set of new formulas, some will be valid, among others, for ChPT and MSSM configurations.

A particularly relevant analytic continuation formula can be obtained for the Lauricella $F_C^{(3)}$ series if one notices that the parameters of all the Lauricella series involved in the series representations of the sunset presented in Section \ref{BERENDSetAl} (in Eqs.(\ref{Berends}), (\ref{BerendsAC}), (\ref{Berends2}) as well as in their symmetrical expressions) admit a common pattern. 

Indeed, owing to the general symmetry property
\begin{align}
F_C^{(3)}(a, b; c, d, e;z_1, z_2, z_3)=F_C^{(3)}(b, a; c, d, e;z_1, z_2, z_3)
\end{align}
one can see that, for appropriate choices of $a$ and $b$, each of the four Lauricella series appearing in Eq.(\ref{Berends}) can be rewritten in the form
\begin{align}
F_C^{(3)}(a, b; c, d, a-b+1;z_1, z_2, z_3)\ ,
\label{form_1}
\end{align} 
each of these appearing in Eq.(\ref{BerendsAC}) can be rewritten as
\begin{align}
F_C^{(3)}\left(a, b; c, d, a-b+1;\frac{z_1}{z_2}, \frac{z_2}{z_3}, \frac{1}{z_3}\right)\ ,
\end{align}
etc.

In fact, it would also be possible to get the following patterns
\begin{align}
F_C^{(3)}(a, b;  a-b+1, d,e;z_1,z_2 , z_3)
\label{form2}
\end{align} 
or
\begin{align}
F_C^{(3)}(a, b; c,  a-b+1, e;z_1, z_2, z_3)
\label{form3}
\end{align}
for all Lauricella series in Eq.(\ref{Berends}), and similar relations for those of Eq.(\ref{BerendsAC}), etc. but due to the invariance properties of the sunset, one will not learn anything new from these other patterns compared to the one of Eq.(\ref{form_1}), therefore we focus now on the latter.

To obtain the analytic continuation of $F_C^{(3)}$ with this particular pattern of its parameters, one starts once more from Eq.(\ref{FC3_F4}) with $e=a-b+1$ but then, instead of Eq.(\ref{F4_AC1}), one uses the following analytic continuation formula derived in \cite{AFGH} for the Appell $F_4$ function
\begin{align}
F_4(\alpha,\beta;\gamma,\alpha-\beta+1;x,y)=M(\alpha,\beta,\gamma;x,y)
	\label{F4_AC2}
\end{align}
where  
\begin{align}
&M\left(\alpha,\beta,\gamma,x,y\right)=(1-y)^{-\alpha}\frac{\Gamma(\alpha-\beta+1)\Gamma(\gamma)}{\Gamma(\alpha)\Gamma(\beta)}\left[\left(\frac{(1+y)^2}{(1-y)^2}\right)^{-\frac{\alpha}{2}}\frac{2^{\alpha-2\beta}}{\sqrt{\pi}}\frac{\Gamma\left(-\beta+\frac{1}{2}\right)\Gamma\left(\frac{1}{2}+\beta\right)}{\Gamma(\alpha-2\beta+1)\Gamma(2\beta-\alpha)}\right.\nonumber\\
&\times N\left(\alpha,\beta,2\beta-\alpha;\beta+\frac{1}{2},\gamma;\frac{x}{2(1-y)}\left(\frac{(1-y)^2}{(1+y)^2}\right)^\frac{1}{2},\frac{(1-y)^2}{4(1+y)^2}\right)\nonumber\\
&+\left(\frac{(1+y)^2}{(1-y)^2}\right)^{-\frac{\alpha}{2}+\beta-\frac{1}{2}}2^{\alpha-1}\sqrt{\pi}\left\{-\left(\frac{2x}{y-1}\left(\frac{(1+y)^2}{(1-y)^2}\right)^{\frac{1}{2}}\right)^{-\beta}\frac{1}{\Gamma(\gamma-\beta)\Gamma(1+\beta-\gamma)}\right.\nonumber\\
&\times N\left(\alpha-\beta+1,1+\beta-\gamma,\beta;\frac{3}{2},\alpha-\beta+1;\frac{1-y}{2x}\left(\frac{(1-y)^2}{(1+y)^2}\right)^\frac{1}{2},\frac{(1-y)^2}{4(1+y)^2}\right)\nonumber\\
&+\left(\frac{2x}{y-1}\left(\frac{(1+y)^2}{(1-y)^2}\right)^{\frac{1}{2}}\right)^{\frac{1}{2}-\beta}\frac{1}{\Gamma\left(\frac{1}{2}+\beta-\gamma\right)\Gamma\left(\frac{1}{2}-\beta+\gamma\right)}\nonumber
\end{align}
\begin{align}
&\left.\left.\times O\left(\alpha-\beta+\frac{1}{2},\beta-\gamma+\frac{1}{2},\beta-\frac{1}{2};\alpha-\beta+\frac{1}{2},\frac{1}{2};\frac{1-y}{2x}\left(\frac{(1-y)^2}{(1+y)^2}\right)^{\frac{1}{2}},\frac{x(1-y)}{2(1+y)^2}\left(\frac{(1+y)^2}{(1-y)^2}\right)^\frac{1}{2}\right)\right\}\right]
	\label{M}
\end{align}
with
\begin{align}
N\left(\alpha,\beta,\gamma;\delta,\lambda;x,y\right)=\sum_{n,p=0}^{\infty}\frac{x^n}{n!}\frac{y^p}{p!}\frac{\Gamma(\alpha+n+2p)\Gamma(\beta+n)\Gamma(\gamma+n)}{\Gamma(\delta+n+p)\Gamma(\lambda+n)}
	\label{N}
\end{align}
and
\begin{align}
O\left(\alpha,\beta,\gamma;\delta,\lambda;x,y\right)=\sum_{n,p=0}^{\infty}\frac{x^n}{n!}\frac{y^p}{p!}\frac{\Gamma(\alpha+n+p)\Gamma(\beta+n-p)\Gamma(\gamma+n-p)}{\Gamma(\delta+n-p)\Gamma(\lambda+n-p)}\ .
	\label{O}
\end{align}
Eq.(\ref{F4_AC2}) is valid in the region
\begin{align}
R\equiv R_0\cap R_1
	\label{R}
\end{align}
where
\begin{align}
R_0\equiv \left\vert \frac{1-y}{2x}\left(\frac{(1-y)^2}{(1+y)^2}\right)^\frac{1}{2}\right\vert+\left\vert \frac{x(1-y)}{2(1+y)^2}\left(\frac{(1+y)^2}{(1-y)^2}\right)^\frac{1}{2}\right\vert<1
\label{white_zone_F4}
\end{align}
and
\begin{align}
R_1\equiv \vert y\vert<1\ .
\label{constraint}
\end{align}

Let us now proceed to the derivation of the analytic continuations of the Lauricella function in Eq.(\ref{form_1}) from the Appell $F_4$ expression above. 

We have
\begin{align}
F_C^{(3)}(a, b; c, d, a-b+1;z_1, z_2, z_3)&=\sum_{m=0}^\infty\frac{(a)_m(b)_m}{(c)_m}\frac{z_1^m}{m!}\ F_4(a+m,b+m;d,a-b+1;z_2,z_3)\ .
	\label{FC3_F4_2}
\end{align}
A careful look at Eqs.(\ref{R}), (\ref{white_zone_F4}) and (\ref{constraint}) shows that Eq.(\ref{F4_AC2}) is valid only for $0<y<1$ \cite{AFGH}. Therefore, we can already guess from Eq.(\ref{FC3_F4_2}) that the analytic continuation of the Lauricella series deduced from Eq.(\ref{F4_AC2}) will be valid only in the $0<z_3<1$ region. The $z_3>1$ region case as well as negative values of $z_3$ will be discussed further.

\subsection{$0<z_3<1$ case\label{z3<1}}

After substituting Eq.(\ref{F4_AC2}) in Eq.(\ref{FC3_F4_2}), we get the following analytic continuation
\begin{align}
F_C^{(3)} (a,b; c,d,a-b+1; z_1,z_2,z_3) = \sum_{i=1}^3T_i\equiv\tilde M(a,b,c,d;z_1,z_2,z_3)
\label{FC3AC}
\end{align}
where we have defined the $\tilde M$ function as the sum of the following three terms
\begin{align}
	T_1 = & \frac{1}{\sqrt{\pi}}\frac{\Gamma(a-b+1) \Gamma(c) \Gamma(d) \Gamma(\frac{1}{2}-b) \Gamma(\frac{1}{2}+b)}{\Gamma(a) \Gamma(b) \Gamma(a-2b+1) \Gamma(2b-a)}  \left(\frac{2}{1-z_3} \left(\frac{(1-z_3)^2}{(1+z_3)^2}\right)^\frac{1}{2}\right)^{a}  2^{-2b} \nonumber \\
	& \times \sum_{m,n,p=0}^\infty \left( \frac{z_1}{2(1-z_3)} \left( \frac{(1-z_3)^2}{(1+z_3)^2} \right)^{\frac{1}{2}} \right)^m \left( \frac{z_2}{2(1-z_3)} \left( \frac{(1-z_3)^2}{(1+z_3)^2} \right)^{\frac{1}{2}} \right)^n \left( \frac{(1-z_3)^2}{4(1+z_3)^2} \right)^p \nonumber \\
	& \quad \times  \frac{1}{m!} \frac{1}{n!} \frac{1}{p!}\frac{ \Gamma(a+m+n+2p) \Gamma(2b-a+m+n)\Gamma(b+m+n)}{ \Gamma(b+m+n+p+\frac{1}{2})\Gamma(c+m) \Gamma(d+n)}\ ,
	\label{T1}
\end{align}
\begin{align}
	T_2 = & - \frac{\sqrt{\pi}}{2} \frac{\Gamma(a-b+1) \Gamma(c) \Gamma(d) }{ \Gamma(a) \Gamma(b) \Gamma(d-b) \Gamma(1+b-d) } \left( \frac{2}{1-z_3} \left( \frac{(1-z_3)^2}{(1+z_3)^2} \right)^{\frac{1}{2}} \right)^a \left( \frac{2z_2}{z_3-1} \left( \frac{(1-z_3)^2}{(1+z_3)^2} \right)^{\frac{1}{2}} \right)^{-b} \nonumber \\
	& \times  \left( \frac{(1-z_3)^2}{(1+z_3)^2} \right)^{\frac{1}{2}}  \sum_{m,n,p=0}^\infty \left( \frac{z_1}{z_2} \right)^m\left( \frac{1-z_3}{2 z_2} \left( \frac{(1-z_3)^2}{(1+z_3)^2} \right)^{\frac{1}{2}} \right)^n  \left( \frac{(1-z_3)^2}{4 (1+z_3)^2} \right)^p \frac{1}{m!} \frac{1}{n!} \frac{1}{p!} \nonumber \\
	& \quad \times \frac{\Gamma(b+m+n) \Gamma(1+b-d+m+n) \Gamma(1+a-b+n+2p) }{\Gamma(c+m) \Gamma(1+a-b+n) \Gamma(\frac{3}{2}+n+p) }
	\label{T2}
\end{align}
and
\begin{align}
	T_3 = & \frac{\sqrt{\pi}}{2} \frac{\Gamma(a-b+1) \Gamma(c) \Gamma(d) }{\Gamma(a) \Gamma(b) \Gamma(\frac{1}{2}-b+d) \Gamma(\frac{1}{2}+b-d)}  \left( \frac{2}{1-z_3} \left( \frac{(1-z_3)^2}{(1+z_3)^2} \right)^{\frac{1}{2}} \right)^a \left( \frac{2z_2}{z_3-1} \left( \frac{(1-z_3)^2}{(1+z_3)^2} \right)^{\frac{1}{2}} \right)^{\frac{1}{2}-b} \nonumber \\
	& \times \sum_{m,n,p=0}^\infty \left( \frac{z_1}{z_2} \right)^m\left( \frac{1-z_3}{2 z_2} \left( \frac{(1-z_3)^2}{(1+z_3)^2} \right)^{\frac{1}{2}} \right)^n  \left(  \frac{z_2(1-z_3)}{2(1+z_3)^2} \left(\frac{(1+z_3)^2}{(1-z_3)^2} \right)^{\frac{1}{2}} \right)^p \frac{1}{m!} \frac{1}{n!} \frac{1}{p!} \nonumber \\
	& \quad \times \frac{\Gamma(b-\frac{1}{2}+m+n-p) \Gamma(\frac{1}{2}+b-d+m+n-p) \Gamma(\frac{1}{2}+a-b+n+p) }{\Gamma(c+m) \Gamma(\frac{1}{2}+n-p) \Gamma(\frac{1}{2}+a-b+n-p) }\ .
	\label{T3}
\end{align}
We now have to determine the convergence regions of each of these three new series.
To determine the convergence region of $T_1$, we note that this series has the same convergence behaviour as the series
\begin{align}
	\sim \sum_{m,n,p=0}^\infty \frac{x^m}{m!} \frac{y^n}{n!} \frac{z^p}{p!} \frac{\Gamma(A+m+n) \Gamma(B+m+n+2p) \Gamma(C+m+n)}{\Gamma(D+m) \Gamma(E+n) \Gamma(F+m+n+p)}
\end{align}
with
\begin{align}
	& x = \frac{z_1}{2(1-z_3)} \left( \frac{(1-z_3)^2}{(1+z_3)^2} \right)^{\frac{1}{2}} \nonumber \\
	& y = \frac{z_2}{2(1-z_3)} \left( \frac{(1-z_3)^2}{(1+z_3)^2} \right)^{\frac{1}{2}} \nonumber \\
	& z = \frac{(1-z_3)^2}{4(1+z_3)^2}\ .
\end{align}
This is due to the fact that the convergence properties of multiple gaussian hypergeometric series are independent of their parameters (excluding exceptional values of the latter) \cite{Srivastava}.

By expressing the sum over $p$ as a ${}_2F_1$ hypergeometric function after using the duplication formula on $\Gamma(B+m+n+2p)$, the above may be written as
\begin{align}
	 \sum_{m,n,p=0}^\infty \frac{x^m}{m!} \frac{y^n}{n!} \frac{\Gamma(A+m+n) \Gamma(B+m+n) \Gamma(C+m+n)}{\Gamma(D+m) \Gamma(E+n) \Gamma(F+m+n)} 
	{}_2F_1 & \left[ \begin{array}{c}
		\frac{B}{2}+\frac{m}{2}+\frac{n}{2}, \frac{1}{2}+\frac{B}{2}+\frac{m}{2}+\frac{n}{2} \\
		F+m+n \\
	\end{array}	\bigg| 4z \right].
	\label{seriesT1}
\end{align}
Now, using the following result from \cite{Erdelyi}
\begin{align}
	{}_2F_1 & \left[ \begin{array}{c}
			\alpha, \alpha + \frac{1}{2} \\
			2 \alpha \\
		\end{array}	\bigg| X \right] = \left( \frac{1}{2}+\frac{1}{2} \sqrt{1-X} \right)^{1-2\alpha} (1-X)^{-\frac{1}{2}} 
		\label{Erdelyi2F1_1}
\end{align}
which is valid for $\vert X\vert<1$, we can write 
\begin{align}
{}_2F_1 & \left[ \begin{array}{c}
	\frac{B}{2}+\frac{m}{2}+\frac{n}{2}, \frac{1}{2}+\frac{B}{2}+\frac{m}{2}+\frac{n}{2} \\
	F+m+n \\
\end{array}	\bigg| 4z \right] = \left( \frac{1}{2}+\frac{1}{2} \sqrt{1-4z} \right)^{1-B-m-n} (1-4z)^{-\frac{1}{2}} 
\label{Erdelyi2F1_2}
\end{align}
if we set $\alpha = \frac{B}{2}+\frac{m}{2}+\frac{n}{2}$ and $F=B$. As already mentioned, the latter choice does not affect the convergence of the underlying series in Eq.(\ref{seriesT1}), which is dependent upon the structure of the summation parameters $m$, $n$ and $p$ rather than on the values of $A$, $B$, etc. The series can then be said to go as
\begin{align}
	\sim \sum_{m,n,p=0}^\infty \frac{r^m}{m!} \frac{s^n}{n!} \frac{\Gamma(A+m+n) \Gamma(C+m+n)}{\Gamma(D+m) \Gamma(E+n)} 
	\label{F4_equiv}
\end{align}
where
\begin{align}
	& r \equiv \frac{2 x}{1+\sqrt{1-4z}} = \frac{z_1}{1-z_3} \left( \frac{(1-z_3)^2}{(1+z_3)^2} \right)^{\frac{1}{2}} \left( 1 + \left( \frac{4 z_3}{(1+z_3)^2} \right)^{\frac{1}{2}} \right)^{-1} \nonumber \\
	& s \equiv \frac{2 y}{1+\sqrt{1-4z}} = \frac{z_2}{1-z_3} \left( \frac{(1-z_3)^2}{(1+z_3)^2} \right)^{\frac{1}{2}} \left( 1 + \left( \frac{4 z_3}{(1+z_3)^2} \right)^{\frac{1}{2}} \right)^{-1}.
\end{align}
The series in Eq.(\ref{F4_equiv}) is an Appell $F_4$ double series whose convergence region is $\sqrt{|r|}+\sqrt{|s|}<1$. It is therefore established that the series in $T_1$ converges in the intersection $\tilde R_1$ of this region with the region $\left\vert\frac{(1-z_3)^2}{(1+z_3)^2}\right\vert<1$ that comes from Eq.(\ref{Erdelyi2F1_2}):  $\tilde R_1\equiv (\sqrt{|r|}+\sqrt{|s|}<1) \cap \left\vert\frac{(1-z_3)^2}{(1+z_3)^2}\right\vert<1$.

We now have to deal with the convergence properties of the series in $T_2$. Its convergence region is determined following the same steps as for the case of $T_1$ above. Finally, one gets $(\tilde R_2\equiv\sqrt{|r|}+\sqrt{|s|} < 1)\cap \left\vert\frac{(1-z_3)^2}{(1+z_3)^2}\right\vert<1$, with
\begin{align}
	& r = \frac{z_1}{z_2} \nonumber \\
	& s = \frac{1-z_3}{z_2} \left( \frac{(1-z_3)^2}{(1+z_3)^2} \right)^{\frac{1}{2}} \left( 1+\left(\frac{4 z_3}{(1+z_3)^2}\right)^{\frac{1}{2}} \right)^{-1}.
\end{align}

The region of convergence of the series in $T_3$ follows from a slightly different derivation, where the use of
\begin{align}
{}_2F_1 & \left[ \begin{array}{c}
	\alpha, \alpha+\frac{1}{2} \\
	\frac{1}{2} \\
\end{array}	\bigg| X \right] = \frac{1}{2}\left(1 + \sqrt{X} \right)^{-2\alpha}+\frac{1}{2}\left(1 - \sqrt{X} \right)^{-2\alpha},
\end{align}
valid for $\vert X\vert<1$, is needed \cite{Erdelyi}. At the end, one obtains the region of convergence of the series in $T_3$ as $\tilde R_3\equiv(\vert r_1\vert +\vert s_1\vert <1)\cap (\vert r_2\vert +\vert s_2\vert <1)\cap \vert\frac{z_1}{z_2}\vert<1$ with
\begin{align}
	& r_1 = \frac{1-z_3}{2z_2}\left(\frac{(1-z_3)^2}{(1+z_3)^2}\right)^{\frac{1}{2}}\left(1+\sqrt{\frac{z_1}{z_2}}\right)^{-2}
	 \nonumber \\
	& s_1 = \frac{z_2(1-z_3)}{2(1+z_3)^2}\left(\frac{(1+z_3)^2}{(1-z_3)^2}\right)^{\frac{1}{2}}\left(1+\sqrt{\frac{z_1}{z_2}}\right)^2
\end{align}
and
\begin{align}
	& r_2 = \frac{1-z_3}{2z_2}\left(\frac{(1-z_3)^2}{(1+z_3)^2}\right)^{\frac{1}{2}}\left(1-\sqrt{\frac{z_1}{z_2}}\right)^{-2} \nonumber \\
	& s_2 = \frac{z_2(1-z_3)}{2(1+z_3)^2}\left(\frac{(1+z_3)^2}{(1-z_3)^2}\right)^{\frac{1}{2}}\left(1-\sqrt{\frac{z_1}{z_2}}\right)^2.
\end{align}

\begin{figure}[hbtp]
\centering
\includegraphics[scale=0.6]{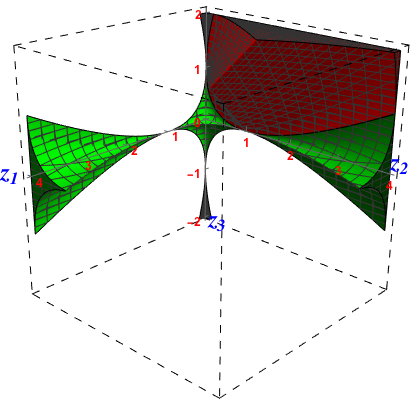}
\caption{Intersection of the convergence regions of Eqs.(\ref{T1}), (\ref{T2}) and (\ref{T3}), in red.}
\label{convInter}
\end{figure}

\begin{figure}[hbtp]
\centering
\includegraphics[scale=0.4]{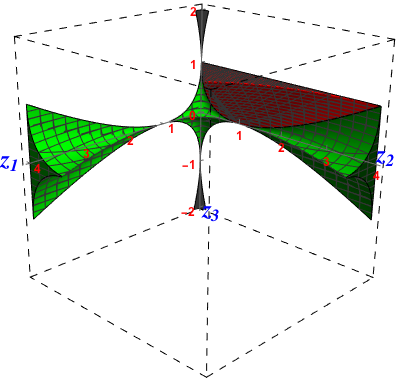}
\includegraphics[scale=0.4]{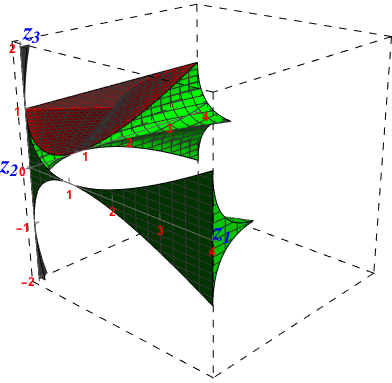}
\caption{\textit{Left}: Region of validity $\tilde R$ of Eqs.(\ref{FC3AC}) and (\ref{S5}), in red. \textit{Right}: same as in left, but shown from another angle of view.}
\label{conv_z3<1}
\end{figure}

The intersection of $\tilde R_1, \tilde R_2$ and $\tilde R_3$ is shown in Fig. \ref{convInter} (red region).
We see, as already guessed just before the present section, that this region is indeed restricted to positive values of $z_3$.
However, the final region of validity of the analytic continuation formula in Eq.(\ref{FC3AC}) must also take into account the constraints coming from the intermediate steps of its derivation, namely those in Eq.(\ref{R}). The $R_0$ constraint does in fact not provide any additional restriction, therefore the main constraint comes from $R_1$ and we conclude that the region of validity of Eq.(\ref{FC3AC}) is the restriction to $0<z_3<1$ of the region plotted in red in Fig. \ref{convInter}. It is explicitly given as 
\begin{align}
\tilde R=\tilde R_1\cap \tilde R_2\cap \tilde R_3\cap (\vert z_3\vert<1)
\label{Rtilde}
\end{align}
and shown in Fig. \ref{conv_z3<1}.

\subsection{$z_3>1$ case\label{z3>1}}

It is easy to derive the $z_3>1$ case from the results shown in the preceding section. To this end, we follow exactly the same steps of derivation as those presented in \cite{AFGH} in order to obtain a new analytic continuation formula for the Appell $F_4(\alpha,\beta;\gamma,\alpha-\beta+1;x,y)$ function valid for $y>1$: in the latter case, we first applied Eq.(\ref{F4_AC1}) on the l.h.s. of Eq.(\ref{F4_AC2}) and then used a slightly rewritten form of the $M(\alpha,\beta,\gamma;x,y)$ function, defined in \cite{AFGH} as $M_M(\alpha,\beta,\gamma;x,y)$, to analytically continue both of the two $F_4$ functions that result.

\begin{figure}[hbtp]
\centering
\includegraphics[scale=0.4]{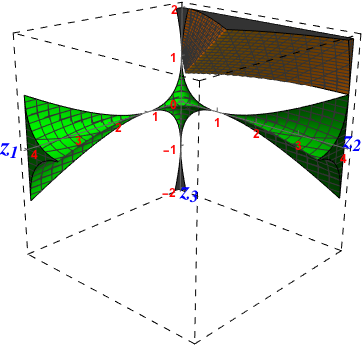}
\includegraphics[scale=0.4]{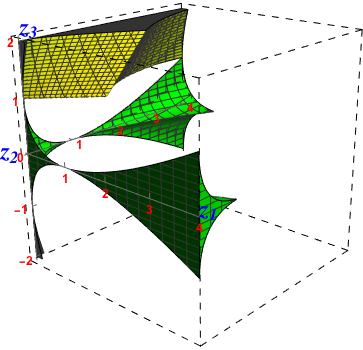}
\caption{\textit{Left}: Region of validity $\tilde R'$ of Eqs.(\ref{FC3AC2}) and (\ref{S6}), in yellow. \textit{Right}: same as in left, but rotated.}
\label{conv_z3>1}
\end{figure}

In the Lauricella case, which generalizes the Appell ones, we obtained in the r.h.s of Eq.(\ref{FC3AC}) an analytic continuation $\tilde M(a,b,c,d;z_1,z_2,z_3)$, valid in the $0<z_3<1$ region, which in fact generalises the $M(\alpha,\beta,\gamma;x,y)$ analytic continuation, valid for $0<y<1$, to the triple series case. Therefore, to obtain a series representation of the Lauricella $F_C^{(3)}$ function valid in the $\vert z_3\vert >1$ region (i.e. the part of the red region of Fig. \ref{convInter} that does not belong to the red region of Fig. \ref{conv_z3<1}), it is sufficient to first apply Eq.(\ref{LauricellaAC}), which is a generalisation of Eq.(\ref{F4_AC1}), on the l.h.s of Eq.(\ref{FC3AC}) and then to apply, on each of the two resulting Lauricella $F_C^{(3)}$ series, the slighlty rewritten form $\tilde M_M(a,b,c,d;z_1,z_2,z_3)$ of the analytic continuation $\tilde M(a,b,c,d;z_1,z_2,z_3)$ where we have performed the following replacements in the overall coefficients of $T_2$ and $T_3$
\begin{align}&\left( \frac{2z_2}{z_3-1}\left( \frac{(1-z_3)^2}{(1+z_3)^2} \right)^{\frac{1}{2}} \right)^{-b}\longmapsto \left( \frac{z_3-1}{2z_2}\left( \frac{(1+z_3)^2}{(1-z_3)^2} \right)^{\frac{1}{2}} \right)^{b}\label{replace1}\\
&\left( \frac{2z_2}{z_3-1} \left( \frac{(1-z_3)^2}{(1+z_3)^2} \right)^{\frac{1}{2}} \right)^{\frac{1}{2}-b} \longmapsto \left( \frac{z_3-1}{2z_2} \left( \frac{(1+z_3)^2}{(1-z_3)^2} \right)^{\frac{1}{2}} \right)^{-\frac{1}{2}+b}.\label{replace2}
\end{align}
This gives the following analytic continuation formula
\begin{align}
F_C^{(3)}(a,b;c,d,a-b+1;z_1,z_2,z_3)=&\frac{\Gamma(a-b+1)\Gamma(b-a)}{\Gamma(b)\Gamma(1-b)}(-z_3)^{-a}\tilde M_M\left(a,b,c,d;\frac{z_1}{z_3}, \frac{z_2}{z_3},\frac{1}{z_3}\right)\nonumber\\
+&\frac{\Gamma(a-b+1)\Gamma(a-b)}{\Gamma(a)\Gamma(a-2b+1)}(-z_3)^{-b}\tilde M_M\left(2b-a,b,c,d;\frac{z_1}{z_3}, \frac{z_2}{z_3},\frac{1}{z_3}\right)
	\label{FC3AC2}
\end{align}
and it is straightforward to show that it is valid in the region
\begin{align}
\tilde R'=\tilde R_1\cap \tilde R_2\cap \tilde R_3\cap (\vert z_3\vert>1)\ ,
\label{Rtilde'}
\end{align}
shown in Fig. \ref{conv_z3>1}. It is therefore just the restriction to $z_3>1$ of the region plotted in red in Fig. \ref{convInter}.

\subsection{Other analytic continuation formulas\label{Other}}

An obvious new analytic continuation formula for the Lauricella series with the specific pattern under study in this paper can be obtained by performing exactly the same analysis as in the preceding sections with the starting point
\begin{align}
F_C^{(3)}(a, b; c, d, a-b+1;z_1, z_2, z_3)&=\sum_{n=0}^\infty\frac{(a)_n(b)_n}{(d)_n}\frac{z_2^n}{n!}\ F_4(a+n,b+n;c,a-b+1;z_1,z_3)
	\label{FC3_F4_3}
\end{align}
instead of Eq.(\ref{FC3_F4_2}).

This is equivalent to use the invariance of the Lauricella series under the exchange of $z_1$ and $z_2$ as well as their corresponding coefficients $c$ and $d$, which preserves the pattern of the other coefficients.

\begin{figure}[hbtp]
\centering
\includegraphics[scale=0.4]{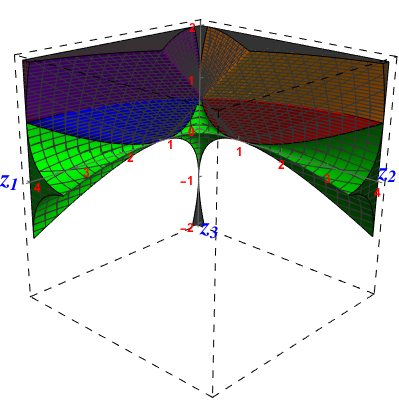}
\includegraphics[scale=0.4,angle=6,origin=C]{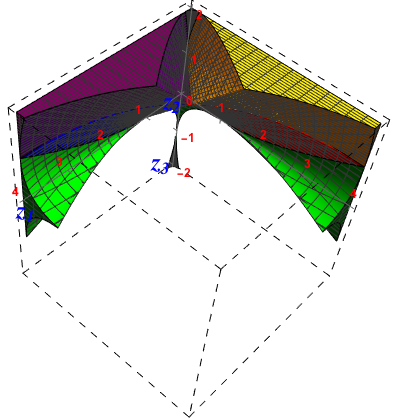}
\caption{\textit{Left:} Regions of validity of Eqs.(\ref{FC3AC3}), in blue, and of Eqs.(\ref{FC3AC4}), in purple. \textit{Right:} view from above.} 
\label{convAC1_z3<1_z3>1}
\end{figure}

One then obtains the two following analytic continuation formulas
\begin{align}
F_C^{(3)}(a,b;c,d,a-b+1;z_1,z_2,z_3)=\tilde M\left(a,b,d,c;z_2, z_1,z_3\right)
	\label{FC3AC3}
\end{align}
which is valid in the region $\tilde R(z_1\leftrightarrow z_2)$ shown in blue in Fig. \ref{convAC1_z3<1_z3>1}, and
\begin{align}
F_C^{(3)}(a,b;c,d,a-b+1;z_1,z_2,z_3)=&\frac{\Gamma(a-b+1)\Gamma(b-a)}{\Gamma(b)\Gamma(1-b)}(-z_3)^{-a}\tilde M_M\left(a,b,d,c;\frac{z_2}{z_3}, \frac{z_1}{z_3},\frac{1}{z_3}\right)\nonumber\\
+&\frac{\Gamma(a-b+1)\Gamma(a-b)}{\Gamma(a)\Gamma(a-2b+1)}(-z_3)^{-b}\tilde M_M\left(2b-a,b,d,c;\frac{z_2}{z_3}, \frac{z_1}{z_3},\frac{1}{z_3}\right)
	\label{FC3AC4}
\end{align}
which is valid in the region $\tilde R'(z_1\leftrightarrow z_2)$ shown in purple in Fig. \ref{convAC1_z3<1_z3>1}.

\section{New series representations for the sunset diagram\label{new}}

We will now use the results presented in Section \ref{ANALYTIC_C} to derive several new series representations of the sunset diagram. 

Let us begin by replacing each of the Lauricella series in Eq.(\ref{Berends}) by their analytic continuation obtained from Eq.(\ref{FC3AC}). We get a new series representation of the sunset diagram $T(p^2, m_1^2, m_2^2, m_3^2, 1, 1, 1)$, valid in the region $\tilde R$, defined in Eq.(\ref{Rtilde}) and shown in Fig. \ref{conv_z3<1}, which can be written as
\begin{align}
&S_5\equiv-m_3^2\left(\frac{m_3^2}{4\pi\mu^2}\right)^{2(\nu-1)}\nonumber \\ 
\times&\left\{
z_1^\nu z_2^\nu \Gamma^2(-\nu)\tilde M( 1+\nu,1, 1+\nu, 1+\nu;z_1, z_2, z_3)\right. \nonumber \\
&-z_1^\nu \Gamma^2(-\nu)\tilde M(1, 1-\nu, 1+\nu,1-\nu; z_1, z_2, z_3) \nonumber \\  
&-z_2^\nu \Gamma^2(-\nu)\tilde M(1, 1-\nu, 1-\nu, 1+\nu; z_1, z_2, z_3) \nonumber \\ 
&\left.-\Gamma(\nu) \Gamma(-\nu)\Gamma(1-2\nu)\tilde M(1-\nu, 1-2\nu,1-\nu, 1-\nu; z_1, z_2, z_3) \right\}.\label{S5}
\end{align}
Now, if instead of working with Eq.(\ref{FC3AC}), one replaces each of the Lauricella series in Eq.(\ref{Berends}) using Eq.(\ref{FC3AC2}), one then obtains

\begin{align}
&S_6\equiv-m_3^2\left(\frac{m_3^2}{4\pi\mu^2}\right)^{2(\nu-1)}\nonumber\\ 
\times &\left\{-\frac{z_1^\nu}{z_3}  \Gamma^2(-\nu)\tilde M_M\left(1,1-\nu;1+\nu,1-\nu;\frac{z_1}{z_3},\frac{z_2}{z_3},\frac{1}{z_3}\right)\right. \nonumber\\
-&\frac{z_2^\nu}{z_3} \Gamma^2(-\nu)\tilde M_M\left(1,1-\nu;1-\nu,1+\nu;\frac{z_1}{z_3},\frac{z_2}{z_3},\frac{1}{z_3}\right)\nonumber\\
-&\frac{(z_1 z_2)^\nu}{z_3} \Gamma^2(-\nu)\tilde M_M\left(1-\nu,1;1+\nu,1+\nu;\frac{z_1}{z_3},\frac{z_2}{z_3},\frac{1}{z_3}\right)\nonumber\\
+&\frac{z_1^\nu}{z_3}(-z_3)^\nu \frac{\Gamma^2(-\nu)\Gamma(\nu)\Gamma(1+\nu)}{\Gamma(2\nu)}\tilde M_M\left(1-2\nu,1-\nu;1+\nu,1-\nu;\frac{z_1}{z_3},\frac{z_2}{z_3},\frac{1}{z_3}\right)\nonumber\\
+&\frac{z_2^\nu}{z_3}(-z_3)^\nu \frac{\Gamma^2(-\nu)\Gamma(\nu)\Gamma(1+\nu)}{\Gamma(2\nu)}\tilde M_M\left(1-2\nu,1-\nu;1-\nu,1+\nu;\frac{z_1}{z_3},\frac{z_2}{z_3},\frac{1}{z_3}\right)\nonumber\\
-&(-z_3)^{\nu-1}\frac{\Gamma^2(-\nu)\Gamma(\nu)\Gamma(1+\nu)}{\Gamma(2\nu)}\tilde M_M\left(1-\nu,1-2\nu;1-\nu,1-\nu;\frac{z_1}{z_3},\frac{z_2}{z_3},\frac{1}{z_3}\right)\nonumber\\
+&\left.(-z_3)^{2\nu-1}\frac{\Gamma^3(\nu)\Gamma(1-2\nu)}{\Gamma(3\nu)}\tilde M_M\left(1-3\nu,1-2\nu;1-\nu,1-\nu;\frac{z_1}{z_3},\frac{z_2}{z_3},\frac{1}{z_3}\right)\right\}\ ,
\label{S6}
\end{align}
which is valid in the $\tilde R'$ region defined in Eq.(\ref{Rtilde'}) and shown in Fig. \ref{conv_z3>1}.

Note that the latter result can of course also be obtained by replacing each of the Lauricella series in Eq.(\ref{BerendsAC}) by their analytic continuation $\tilde M_M(a,b,c,d;z_1,z_2,z_3)$, that we recall to be defined from $\tilde M(a,b,c,d;z_1,z_2,z_3)$ using Eqs.(\ref{replace1}) and (\ref{replace2}).

It is possible to derive more series representations of the sunset using the analytic continuation formulas presented so far. Indeed, due to the fact that the Lauricella $F_C^{(3)}$ series is invariant under any permutation of its variables (and associated parameters) and from the specific form of the parameters of the Lauricella series involved in the sunset representations $S_1$ and $S_2$, that allow us to use our analytic continuation formulas for any of the possible permutations, one can for instance derive the following series
\begin{align}
S_7=-m_3^2\left(\frac{m_3^2}{4\pi\mu^2}\right)^{2(\nu-1)} 
&\left\{
z_1^\nu z_2^\nu \Gamma^2(-\nu)\tilde M( 1+\nu,1, 1+\nu, 1+\nu;z_1, z_3, z_2)\right. \nonumber \\
&-z_1^\nu \Gamma^2(-\nu)\tilde M(1-\nu, 1, 1+\nu,1+\nu; z_1, z_3, z_2) \nonumber \\  
&-z_2^\nu \Gamma^2(-\nu)\tilde M(1, 1-\nu, 1-\nu, 1+\nu; z_1, z_3, z_2) \nonumber \\ 
&\left.-\Gamma(\nu) \Gamma(-\nu)\Gamma(1-2\nu)\tilde M(1-2\nu, 1-\nu,1-\nu, 1+\nu; z_1, z_3, z_2) \right\}\ ,
\end{align}
which is valid in the region $\tilde R(z_2\leftrightarrow z_3)$. 

$S_7$ cannot be obtained from any other series representations of the sunset shown in this paper by using the permutation of masses invariance property of the sunset diagram. From the phenomenological point of view, $S_7$ is quite efficient because one can use it at the same time for the kaon and eta cases of Chiral Perturbation Theory, as well as for all the $``$heavy" sunsets of MSSM considered in the next section. Moreover, $S_7$ is also able to reach negative values of $z_3$ ($p^2<0$) which is not the case for $S_5$ and $S_6$.

Three other series representations of the sunset ($S_8$, $S_9$ and $S_{10}$), derived using the invariance of the Lauricella series under the permutation of some of their variables (and which cannot be obtained from the others series $S_i$ ($i=1,...,7$) by permutations of the masses) are given in the Appendix with their respective regions of validity, see Eqs.(\ref{S8})-(\ref{S10}). $S_7$ ($S_8$) is obtained from analytically continuing $S_1$ after using the permutation of the variables $z_2$ ($z_1$) and $z_3$, while $S_9$ ($S_{10}$) is derived from $S_2$ after permutating the variables $\frac{z_1}{z_3}$ ($\frac{z_2}{z_3}$) and $\frac{1}{z_3}$ (as well as permuting the corresponding parameters). These 4 permutations of variables are the only non trivial ones in the set of the 10 possible permutations. The 6 others permutations give series representations that can be obtained from suitable permutations of the masses in the $S_i$ ($i=5,...,10$).

\begin{figure}[hbtp]
\centering
\includegraphics[scale=0.4]{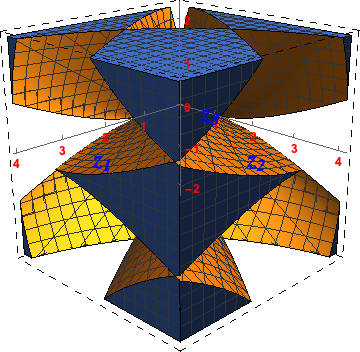}
\caption{New area of the $(z_1,z_2,z_3)$-space that can be reached by adding the regions of validity of the new series representations $S_i$ ($i=5,...,10$) and of the series deduced from them by permutations of the masses.
\label{FinalConv}}
\end{figure}

Combining the regions of validity of our new series representations for the sunset with 4 mass scales, a large new area of the $(z_1,z_2,z_3)$-space is now reachable (see Fig. \ref{FinalConv} and compare with Fig. \ref{Reg2} which combines the  regions of convergence of the 2 series representations given in\cite{Berends:1993ee} as well as those of the series obtained from them by permuting the variables).

\section{$\epsilon$-expansion\label{eps}}

The expressions for the sunset $S_1$ to $S_{10}$ in this work are functions of the small parameter $\epsilon$, which arises from dimensional regularization.  In this section, following the method of \cite{Greynat:2013hox,Greynat:2013zqa}, we show how to obtain their $\epsilon$-expansion about $\epsilon=0$ by using as an example the second term of $S_7$
\begin{align}
    m_3^2 z_1^\nu \left( \frac{m_3^2}{4\pi\mu^2} \right)^{2(\nu-1)}  \Gamma(-\nu)^2 \Tilde{M} (1-\nu,1,1+\nu,1+\nu; z_1, z_3, z_2)
    \label{Eq:S9t2}
\end{align}
where
\begin{align}
    \Tilde{M} (1-\nu,1,1+\nu,1+\nu; z_1, z_3, z_2) = T_1  ( ... ) + T_2  (...) + T_3  (...)\ .
\end{align}
The ellipsis in the arguments of the $T_i$ imply that they are equivalent to the argument of the $\Tilde{M}$ on the LHS of the equation.

$T_1 (1-\nu,1,1+\nu,1+\nu; z_1, z_3, z_2) $, after combining with the factors preceding the $\Tilde{M}$ of Eq.(\ref{Eq:S9t2}) and simplifying it, can be expressed as
\begin{align}
    & T_{1, \epsilon} \equiv \frac{1}{2} \frac{\Gamma (\epsilon )^2}{1-\epsilon} \left( \frac{2}{1-z_3} \sqrt{\frac{(1-z_3)^2}{(1+z_3)^2}} \right)^{\epsilon} \sum_{m,n,p=0}^\infty \bigg\{ \frac{(1)_{m+n} (2-\epsilon)_{m+n} (\epsilon)_{m+n+2p}} {m! n! p! (\tfrac{3}{2})_{m+n+p} (2-\epsilon)_m (2-\epsilon)_n}  \nonumber \\
    & \times
    \left( \frac{z_1}{2(1-z_3)} \sqrt{\frac{(1-z_3)^2}{(1+z_3)^2}} \right)^m
    \left( \frac{z_2}{2(1-z_3)} \sqrt{\frac{(1-z_3)^2}{(1+z_3)^2}} \right)^n
    \left( \frac{(1-z_3)^2}{4(1+z_3)^2} \right)^p \bigg\}\ .
    \label{Eq:S9STT1}
\end{align}

The $\epsilon$-expansion of the above involves expanding the $\epsilon$ dependent part by means of the generalized product rule
\begin{align}
    \prod_{i=1}^n f_i(\epsilon) 
    &= \sum_{k=0}^\infty \frac{\epsilon^k}{k!} \left( \prod_{i=1}^n f_i \right)^{(k)} \nonumber \\
    &= \sum_{k=0}^\infty \frac{\epsilon^k}{k!} \sum_{j_1=0}^k \sum_{j_2=0}^{(k-j_1)} ... \sum_{j_{n-1}=0}^{(k-j_1 ... - j_{n-2})} \binom{k}{j_1,j_2,...,j_n} \prod_{i=1}^{n} f_i^{(j_i)}
\label{Eq:ProductRule}
\end{align}
where
\begin{align*}
    \binom{k}{j_1,j_2,...,j_n} = \frac{k!}{j_1! j_2! ... j_n!}
\end{align*}
is the multinomial coefficient, and $ j_n = k-j_1...-j_{n-1} $.

The derivatives of the Pochhammer, the inverse Pochhammer, and the quotient of two Pochhammer  functions, are given in \cite{Greynat:2013hox,Greynat:2013zqa}:
\begin{align}
\mathcal{P}_m^{(k)} (\alpha) &\equiv \frac{1}{k!} \frac{\partial^k}{\partial \alpha^k} (\alpha)_m \nonumber \\
        & =\begin{cases}
              0 \quad &\text{ for } k>m \\
			(-1)^{m-k} \sum_{l=0}^{m-k} (-1)^l {{m}\choose{l}} s(m-l,k) (\alpha)_l \quad &\text{ for } k \leq m
        \end{cases}
\\
\mathcal{Q}_m^{(k)} (\beta) &\equiv \frac{1}{k!} \frac{\partial^k}{\partial \beta^k} \frac{1}{(\beta)_m}
\nonumber \\
&   =\begin{cases}
        \delta_{k,m} \quad &\text{ for } m=0 \\
		1/(\beta)_m \quad &\text{ for } k=0 \\					
		(-1)^{k+l} \sum_{l=0}^{m-1} \left( l!(m-1-l)! (\beta+l)^{k+1} \right)^{-1} \quad &\text{ for } m,k >0
	\end{cases}
\end{align}

\begin{align}
\mathcal{R}_{m,n}^{(k)} (\alpha,\beta) &\equiv \frac{1}{k!} \frac{\partial^k}{\partial \epsilon^k} \frac{(\alpha)_m}{(\beta)_n}
\nonumber \\
    &   = (-b)^k \left( \delta_{m,n} \delta_{k,0} \left( \frac{a}{b} \right)^n 
            + \sum_{j=0}^{n-1} \frac{r_{j,m,n}(\alpha,\beta)}{(\beta+j)^{k+1}} \right) 
            \quad &\text{ for } m \leq n
\end{align}
where
\begin{align}
	\alpha = A + a \epsilon ,\qquad \beta_j = B + b \epsilon
\end{align}
and
\begin{align}
    r_{j,m,n} (\alpha,\beta) = \frac{(-1)^j \left( A - \tfrac{a}{b} \left( B+j \right) \right)_m }{j! (n-1-j)!}\ .
\end{align}

To be able to use the above formulae for the derivatives, it is necessary to re-express the Pochhammer functions of Eq.(\ref{Eq:S9STT1}) in the following manner
\begin{align}
    & \frac{z_1 m_3^2}{2} \frac{1}{\epsilon^2} \frac{1}{1-\epsilon} 
    \left( \frac{2}{z_1(1-z_2)} \sqrt{\frac{(1-z_2)^2}{(1+z_2)^2}} \left( \frac{4\pi\mu^2}{m_3^2} \right)^2 \right)^\epsilon \Gamma(1+\epsilon)^2 \nonumber \\
    & \times \sum_{m,n,p=0}^\infty \bigg\{
    \left( \frac{z_1}{2(1-z_2)} \sqrt{\frac{(1-z_2)^2}{(1+z_2)^2}} \right)^m
    \left( \frac{z_3}{2(1-z_2)} \sqrt{\frac{(1-z_2)^2}{(1+z_2)^2}} \right)^n
    \left( \frac{(1-z_2)^2}{4(1+z_2)^2} \right)^p \nonumber \\
    & \quad \times \frac{1}{m!} \frac{1}{n!} \frac{1}{p!}
    \frac{(1)_{m+n}}{(\tfrac{3}{2})_{m+n+p}} (\epsilon)_{m+n+2p} \frac{(2-\epsilon+m)_n}{(2-\epsilon)_n} \bigg\}\ .
\end{align}

Applying Eq.(\ref{Eq:ProductRule}) to the above, we get the expansion of Eq.(\ref{Eq:S9STT1}) about $\epsilon=0$
\begin{align}
    & T_{1, \epsilon} \equiv \frac{m_1^2}{2} \sum_{m,n,p=0}^\infty \bigg\{
    \left( \frac{z_1}{2(1-z_2)} \sqrt{\frac{(1-z_2)^2}{(1+z_2)^2}} \right)^m
    \left( \frac{z_3}{2(1-z_2)} \sqrt{\frac{(1-z_2)^2}{(1+z_2)^2}} \right)^n \nonumber \\
    & \times \left( \frac{(1-z_2)^2}{4(1+z_2)^2} \right)^p \frac{1}{m!} \frac{1}{n!} \frac{1}{p!}
    \frac{(1)_{m+n}}{(\tfrac{3}{2})_{m+n+p}} \sum_{k=0}^\infty \epsilon^{k-2} \sum_{k_1=0}^k \sum_{k_2=0}^{(k-k_1)}  \sum_{k_3=0}^{(k-k_1-k_2)} 
    \nonumber \\
    & \quad \times \bigg[ F_1(k_1) F_2(k_2)\mathcal{P}_{m+n+2p}^{(k_3)} (\epsilon)
    \mathcal{R}_{n,n}^{(k-k_1-k_2-k_3)} (2-\epsilon+m,2-\epsilon) \bigg]_{\epsilon=0} \bigg\}
\end{align}

where
\begin{align}
    u &\equiv \frac{2}{z_1(1-z_2)} \sqrt{\frac{(1-z_2)^2}{(1+z_2)^2}} \left( \frac{4\pi\mu^2}{m_3^2} \right)^2,
\\
    F_1(k) &\equiv \frac{1}{k!} \frac{\partial^k}{\partial \epsilon^k}
\frac{u^\epsilon}{1-\epsilon}  = \left( \frac{1}{(1-\epsilon)^{k+1}} +
\sum_{i=0}^{k-1} \frac{1}{(k-i)!}
\frac{\log^{k-i}(u)}{(1-\epsilon)^{i+1}} \right) u^\epsilon,
\\
    F_2(k) &\equiv \frac{1}{k!}\frac{\partial^k}{\partial \epsilon^k} \Gamma(1+\epsilon)^2.
\end{align}
A similar procedure yields the expansion of
\begin{align}
 T_{2, \epsilon} \equiv m_3^2 z_1^\nu \left( \frac{m_3^2}{4\pi\mu^2} \right)^{2(\nu-1)}  \Gamma(-\nu)^2 T_2 (1-\nu,1,1+\nu,1+\nu; z_1, z_3, z_2)
 \end{align}
as
\begin{align}
     & T_{2, \epsilon} = \frac{m_1^2}{2} \left(
\frac{z_3}{1-z_2} \sqrt{\frac{(1-z_2)^2}{(1+z_2)^2}} \right)^{-1}
\left( \frac{(1-z_2)^2}{(1+z_2)^2} \right)^{\tfrac{1}{2}}
\sum_{m,n,p=0}^\infty \Bigg\{ \frac{1}{m!} \frac{1}{n!} \frac{1}{p!}
\frac{(1)_{m+n}}{(\tfrac{3}{2})_{n+p}} \nonumber \\
     & \times \left( \frac{z_1}{z_3} \right)^m
    \left( \frac{1-z_2}{2z_3} \sqrt{\frac{(1-z_2)^2}{(1+z_2)^2}}
\right)^n  \left( \frac{(1-z_2)^2}{4(1+z_2)^2} \right)^p
\sum_{k=0}^\infty \epsilon^{k-2} \sum_{k_1=0}^k \sum_{k_2=0}^{(k-k_1)}
 \sum_{k_3=0}^{(k-k_1-k_2)}  \nonumber \\
    & \quad \times \bigg[  F_1(k_1) F_2(k_2)
\mathcal{P}_{n+2p}^{(k_3)} (\epsilon)
    \mathcal{R}_{m,m}^{(k-k_1-k_2-k_3)} (\epsilon+n,2-\epsilon)
\bigg]_{\epsilon=0} \Bigg\}
\end{align}
and of 
\begin{align}
T_{3, \epsilon} \equiv m_3^2 z_1^\nu \left( \frac{m_3^2}{4\pi\mu^2} \right)^{2(\nu-1)}  \Gamma(-\nu)^2 T_3 (1-\nu,1,1+\nu,1+\nu; z_1, z_3, z_2)
\end{align}
as
\begin{align}
     & T_{3, \epsilon} = \frac{m_1^2 \sqrt{\pi}}{2} \left( \frac{2 z_3}{z_2-1} \sqrt{\frac{(1-z_2)^2}{(1+z_2)^2}} \right)^{-\tfrac{1}{2}} \sum_{m,n,p=0}^\infty \Bigg\{ \frac{1}{m!} \frac{1}{n!} \frac{1}{p!} \frac{(\tfrac{1}{2})_{m+n-p}}{(\tfrac{1}{2})_{n-p}} \left( \frac{z_1}{z_3} \right)^m \nonumber \\
     & \times 
    \left( \frac{1-z_2}{2z_3} \sqrt{\frac{(1-z_2)^2}{(1+z_2)^2}} \right)^n  \left( \frac{z_3(1-z_2)}{2(1+z_2)^2} \sqrt{\frac{(1+z_2)^2}{(1-z_2)^2}} \right)^p \sum_{k=0}^\infty \epsilon^{k-2} \sum_{k_1=0}^k \sum_{k_2=0}^{(k-k_1)}  \sum_{k_3=0}^{(k-k_1-k_2)}  \nonumber \\
    & \quad \times \bigg[  F_3(k_1) F_4(k_2) \mathcal{P}_{n+p}^{(k_3)} (\epsilon-\tfrac{1}{2})
    \mathcal{R}_{m,m}^{(k-k_1-k_2-k_3)} (-\tfrac{1}{2}+\epsilon+n-p, 2-\epsilon) \bigg]_{\epsilon=0} \Bigg\}
\end{align}
where
\begin{align}
    F_3(k) &\equiv \frac{1}{k!} \frac{\partial^k}{\partial \epsilon^k}
\frac{u^\epsilon}{(\epsilon-1)^2}  = \left( \frac{(-1)^k
(k+1)}{(\epsilon-1)^{k+2}} + \sum_{i=0}^{k-1} \frac{(-1)^i (i+1)
\log^{k-i}(u)}{(k-i)! (\epsilon-1)^{i+2}} \right) u^\epsilon,
\\
    F_4(k) &\equiv \frac{1}{k!}\frac{\partial^k}{\partial \epsilon^k} \frac{\Gamma(1+\epsilon)^2 \Gamma(2-\epsilon)^2}{\Gamma(\tfrac{3}{2}-\epsilon)}\ .
\end{align}

The full $\epsilon$ expansion of Eq.(\ref{Eq:S9t2}) is therefore given by $T_{1, \epsilon} + T_{2, \epsilon}  + T_{3, \epsilon}$.  The above exercise can be carried out for the other terms of $S_7$ and the other sunsets representations, as required by physical considerations.

In this section, $S_7$ was used as a representative sunset as it covers the $``$heavy" MSSM and the kaon and eta ChPT cases considered in the next section.

\section{Some physical applications\label{physics}}

In this section, we show two situations where the analytic continuation formulas of Section \ref{new} and of the Appendix can be used. In each case, checks have been performed by a numerical comparison between our new series representations of the sunset diagram and the numerical integration of the corresponding Feynman parameter integral representations (choosing a finite value for $\epsilon$).

\subsection{Chiral perturbation theory}

The values $m_1=m_2=m_K$, $m_3=m_\eta$  (or any permutation of them) and $p^2=m_\pi^2$ define the pion sunset configuration in ChPT, whereas $m_1=m_\pi$, $m_2=m_K$, $m_3=m_\eta$ (or any permutation of them) and $p^2=m_K^2$ define the kaon one, and $m_1=m_\pi$, $m_2=m_3=m_K$ (or any permutation of them) and $p^2=m_\eta^2$ the eta one.

These sunsets diagrams, which depend on three different masses and which have been studied in \cite{Ananthanarayan:2017qmx, Ananthanarayan:2018irl, Ananthanarayan:2016pos, Ananthanarayan:2017yhz}, were the last missing pieces in the analytic expressions of the masses and decay constants of the light pseudoscalars at two-loop level in ChPT.

In Section \ref{BERENDSetAl}, we recalled the well-known result for the general four mass scale sunset $T_{123}$ that converges for $\sqrt{|z_1|}+\sqrt{|z_2|}+\sqrt{|z_3|}<1$, and which is given as a linear combination of the Lauricella $F_C^{(3)}$ series (see Eq.(16) of \cite{Berends:1993ee} or Eq.(\ref{Berends}) in the present paper). Its analytic continuation to the region $\sqrt{|z_3|} > \sqrt{|z_1|}+\sqrt{|z_2|}+1$ is given in Eq.(24) of \cite{Berends:1993ee} (see also Eq.(\ref{BerendsAC}) in the present paper, where some misprints have been corrected), again in terms of the $F_C^{(3)}$ function. The authors of \cite{Berends:1993ee} stated that provided that $m_1+m_2<m_3$, one could extend the region of convergence of Eq.(\ref{Berends}) to $\vert p^2\vert<(m_1+m_2+m_3)^2$, with careful numerical treatment in the annulus $(m_1+m_2-m_3)^2<\vert p^2\vert<(m_1+m_2+m_3)^2$.

As already discussed in the introduction, with $z_1 \equiv m_1^2/m_3^2$, $z_2 \equiv m_2^2/m_3^2$ and $z_3 \equiv p^2/m_3^2$, neither of the above cases corresponds to the physical kaon, eta or pion sunset configurations. This is true under any exchange of the masses in the propagators, which can be shown to leave the $T_{123}$ integral invariant. This is due to the fact that although the chiral perturbation theory sunsets do satisfy $\vert p^2\vert<(m_1+m_2+m_3)^2$, they always violate the $m_1+m_2<m_3$ condition, as well as $m_1+m_2+\sqrt{\vert p^2\vert}<m_3$ and $m_1+m_2+m_3<\sqrt{\vert p^2\vert}$, for any of the permutations of the masses.

In fact, it is easy to check that the pion configuration, with the choice of masses above, does fall into the region of validity of $S_8$ which can then be used to evaluate it. This series is the only one that can be used for the pion case, even after permutations of the masses.

The kaon configuration does fall into the regions of validity of two of our new series which do have a non-empty intersection: $S_5$ and $S_7$. One can therefore choose either one or the other to evaluate it.

In passing, although the particular set of masses and squared external momentum defined above for the kaon configuation $m_1=m_\pi$, $m_2=m_K$, $m_3=m_\eta$ and $p^2=m_K^2$ does not belong to the region of validity of $S_{10}$ defined after Eq.(\ref{S10}), the numerical evaluation of the latter for this particular set is in fact in perfect agreement with the result coming from the corresponding Feynman parameter integral representation. This is due to the fact that this configuration of masses is exactly on the boundary $\frac{z_3}{z_2}=1$ of the region of validity of $S_{10}$ where the latter is in fact also valid. In Sections  \ref{z3<1} and \ref{z3>1}, we have performed an analysis for $0<z_3<1$ and for $z_3>1$, but we did not try to prove anything at $z_3=1$. The reasons are that on one hand results are in general harder to prove there and on the other hand, as we have seen in the kaon example, it is not necessary to use $S_{10}$ on its boundary because $S_5$ and $S_7$ can very well do the job. 
As a last remark, by appropriate permutations of the masses, it is also possible to use $S_6$ or $S_9$, on their boundary, to evaluate the kaon configuration.

The eta configuration with $m_1=m_\pi$, $m_2=m_3=m_K$ and $p^2=m_\eta^2$ does fall into the regions of validity of $S_6$ and $S_{10}$ and on the boundary of $S_7$. By an appropriate permutation of the masses, it can be shown that $S_9$ can also be used to evaluate the eta configuration.

\subsection{Minimal Supersymmetric Standard Model\label{MSSM}}

In the recent work \cite{Borowka:2018anu}, the full 2-loop QCD contributions to the lightest Higgs-boson mass in the MSSM with complex parameters have been computed. This implies sunset diagrams evaluations, which the authors of \cite{Borowka:2018anu} computed, as well as other 2-loop self energy master integrals, using numerical packages such as $\mathsf{SecDec}$ \cite{SecDec} (we refer the reader to \cite{Borowka:2018anu} for details). Among the MI considered in \cite{Borowka:2018anu}, those of the sunset type are the following 24 different configurations: for $p^2=m^2_{H^+}$ or $p^2=m^2_W$ on the external legs, the possible corresponding internal masses are given by the triplets $(m_{\tilde b_1}, m_t, m_{\tilde g})$, $(m_{\tilde b_2}, m_t, m_{\tilde g})$, $(m_{\tilde t_1}, m_b, m_{\tilde g})$ and $(m_{\tilde t_2}, m_b, m_{\tilde g})$. For $p^2=m^2_{h_1}$, $p^2=m^2_{h_2}$, $p^2=m^2_{h_3}$ or $p^2=m^2_Z$, the triplets are $(m_{\tilde b_1}, m_b, m_{\tilde g})$, $(m_{\tilde b_2}, m_b, m_{\tilde g})$, $(m_{\tilde t_1}, m_t, m_{\tilde g})$ and $(m_{\tilde t_2}, m_t, m_{\tilde g})$.

A particular parameter point
used in \cite{Borowka:2018anu} is\footnote{We thank the authors of \cite{Borowka:2018anu} for providing us these numbers as well as the list of their sunset MI.} 
\begin{align}
\label{parameter_point}
& m_{b}=2.47211388094981\;\textrm{GeV}\nonumber\\
& m_{\tilde b_1}=1999.63574080077\;\textrm{GeV}\nonumber\\
& m_{\tilde b_2}=2101.33691741576\;\textrm{GeV}\nonumber\\
& m_{t}=173.2\;\textrm{GeV}\nonumber\\
& m_{\tilde t_1}=1933.51658629971\;\textrm{GeV}\nonumber\\
& m_{\tilde t_2}=2174.24984231707\;\textrm{GeV}\nonumber\\
& m_{\tilde{g}}=2500\;\textrm{GeV}\\ 
& m^2_W = 6461.748225\;\textrm{GeV}\nonumber\\
& m^2_Z = 8315.17839376\;\textrm{GeV}\nonumber\\
& m_{h_1}^2=15771.9521491081\;\textrm{GeV}\nonumber
\end{align}
\begin{align}
& m_{h_2}^2=2.24109098708842\times 10^6\;\textrm{GeV}\nonumber\\
& m_{h_3}^2=2.24109388932544\times 10^6\;\textrm{GeV}\nonumber\\
& m^2_{H^+}=2.25\times10^6\;\textrm{GeV}\nonumber
\end{align}
Using the numerical values of this particular parameter point, it is easy to check that all the $``$heavy" sunsets (\textit{i.e} those with $p^2=m^2_{H^+}$, $p^2=m^2_{h_2}$ and $p^2=m^2_{h_3}$) cannot be computed from series others than $S_5$ or $S_7$, whereas the $``$light" ones do fall into the convergence region of $S_1$.

It is interesting to note that $S_7$ is able to handle the kaon and eta cases in ChPT as well as all the $``$heavy" configurations of MSSM.

\section{Conclusions and outlook}

In this work, we have provided new series representations $S_i$ ($i=5,...,10$) for the two-loop sunset diagram with four mass scales, following an analytic continuation procedure that we have applied to an old series representation of the sunset diagram given in \cite{Berends:1993ee} in terms of Lauricella $F_C^{(3)}$ series.  The $S_i$ ($i=5,...,10$) are obtained, in our analytic continuation procedure, by the application of non-trivial analytic continuation formulas derived in \cite{AFGH} for the Appell $F_4$ function. Basing ourself on the example of $S_7$, we then have shown how the $\epsilon$-expansion of the $S_i$ can be performed, at any order in $\epsilon$.

These new series representations  give access to large regions of the mass scale parameters space that were unreachable with the representations presented in \cite{Berends:1993ee}.
It may be recalled that
the new regions encompass realistic physical situations, including chiral perturbation
theory and related investigations on the lattice.   Furthermore, in extensions of the SM
such as the MSSM, one also encounters situations where the original series representations of \cite{Berends:1993ee}
fail to converge but where we have shown that ours can be used.   
Although the subject is by now quite classical, the present investigation
recalls that one must have a handle on rigorous analysis of convergence properties, analytic continuation, of even relatively simple
configurations arising in field theory. In this context, although Lauricella functions are known for a long time, the mathematical study of their analytic continuations in terms of series representations is still, to our knowledge, far from complete (see \cite{Exton1} for some results). Since these objects appear in varied contexts of the physical science, from experimental physics in the description of the solid angle subtended at a disk source by a non-coaxial parallel-disk detector \cite{FriotRuby}, to string theory, in the expression of exact string scattering amplitudes \cite{Lai:2016ekc}, in passing by general relativity, in the solutions of the equations of motion of test particle and photon in Kerr and Kerr-(anti) de Sitter spacetimes \cite{Kraniotis:2006ux} and in many other situations, it would be interesting to try to fill this gap. Motivated by quantum field theory, the present investigation is a first step in this direction for the $F_C^{(3)}$ case, the generalisation to situations where $d$ is not restricted to be equal to $a-b+1$ being under way. Concerning the sunset diagram in particular, the exploration of the regions not covered by Fig.\ref{FinalConv} and Fig.\ref{Reg2}, as well as the $L$-loop generalisation, are left for future investigations.

\vspace{1cm}

{\bf Acknowledgements}

B.A. acknowledges partial support from the MSIL Chair of the Division of Physical and Mathematical
Sciences, Indian Institute of Science during the course of this work.  S.G. thanks Ulf-G. Meissner for supporting the research through grants. S.F. thanks Centre for High Energy Physics, Indian Institute of Science of Bangalore and S.G. thanks Institut de Physique Nucléaire d'Orsay, Université Paris-Sud for their hospitality during the course of this work.
S.F. and S.G would also like to thank the \textit{Les Houches School of Physics} organizers, Sacha Davidson and others, for their hospitality, since the first studies on some subjects linked to this work started there. 

\section*{Appendix\label{Appendix}}

We give here three alternative series representations of the sunset diagram $T(p^2, m_1^2, m_2^2, m_3^2, 1, 1, 1)$ that have not already been presented explicitly in the main body of this paper. These are obtained thanks to the invariance of the Lauricella series under permutations of some of their variables (see Section \ref{new} for details). They cannot be derived from the $S_i$ ($i=5,6,7$) using the invariance of the sunset under permutations of the masses. We also give their corresponding region of validity. 

\begin{align}
S_8=-m_3^2\left(\frac{m_3^2}{4\pi\mu^2}\right)^{2(\nu-1)} 
&\left\{
z_1^\nu z_2^\nu \Gamma^2(-\nu)\tilde M( 1+\nu,1, 1+\nu, 1+\nu;z_3, z_2, z_1)\right. \nonumber \\
&-z_1^\nu \Gamma^2(-\nu)\tilde M(1, 1-\nu, 1+\nu,1-\nu; z_3, z_2, z_1) \nonumber \\  
&-z_2^\nu \Gamma^2(-\nu)\tilde M(1-\nu, 1, 1+\nu, 1+\nu; z_3, z_2, z_1) \nonumber \\ 
&\left.-\Gamma(\nu) \Gamma(-\nu)\Gamma(1-2\nu)\tilde M(1-2\nu, 1-\nu,1+\nu, 1-\nu; z_3, z_2, z_1) \right\}
\label{S8}
\end{align}
is valid in the region $\tilde R(z_1\leftrightarrow z_3)$.

\begin{align}
S_{9}=& -m_3^2\left(\frac{m_3^2}{4\pi\mu^2}\right)^{2(\nu-1)}\nonumber\\ 
\times &\left\{-\frac{z_1^\nu}{z_3} \Gamma^2(-\nu)\tilde M_M\left(1,1-\nu,1+\nu,1-\nu;\frac{1}{z_3},\frac{z_2}{z_3},\frac{z_1}{z_3}\right)\right. \nonumber\\
-&\frac{z_2^\nu}{z_3}\Gamma^2(-\nu)\tilde M_M\left(1-\nu,1,1+\nu,1+\nu;\frac{1}{z_3},\frac{z_2}{z_3},\frac{z_1}{z_3}\right)\nonumber\\
-&\frac{(z_1 z_2)^\nu}{z_3}  \Gamma^2(-\nu)\tilde M_M\left(1,1-\nu,1-\nu,1+\nu;\frac{1}{z_3},\frac{z_2}{z_3},\frac{z_1}{z_3}\right)\nonumber\\
+&\frac{z_1^\nu}{z_3}(-z_3)^\nu \frac{\Gamma^2(-\nu)\Gamma(\nu)\Gamma(1+\nu)}{\Gamma(2\nu)}\tilde M_M\left(1-\nu,1-2\nu,1-\nu,1-\nu;\frac{1}{z_3},\frac{z_2}{z_3},\frac{z_1}{z_3}\right)\nonumber\\
+&\frac{z_2^\nu}{z_3}(-z_3)^\nu \frac{\Gamma^2(-\nu)\Gamma(\nu)\Gamma(1+\nu)}{\Gamma(2\nu)}\tilde M_M\left(1-2\nu,1-\nu,1-\nu,1+\nu;\frac{1}{z_3},\frac{z_2}{z_3},\frac{z_1}{z_3}\right)\nonumber\\
-&(-z_3)^{\nu-1}\frac{\Gamma^2(-\nu)\Gamma(\nu)\Gamma(1+\nu)}{\Gamma(2\nu)}\tilde M_M\left(1-2\nu,1-\nu,1+\nu,1-\nu;\frac{1}{z_3},\frac{z_2}{z_3},\frac{z_1}{z_3}\right)\nonumber\\
+&\left.(-z_3)^{2\nu-1}\frac{\Gamma^3(\nu)\Gamma(1-2\nu)}{\Gamma(3\nu)}\tilde M_M\left(1-3\nu,1-2\nu,1-\nu,1-\nu;\frac{1}{z_3},\frac{z_2}{z_3},\frac{z_1}{z_3}\right)\right\}
\label{S9}
\end{align}
is valid in the region $\tilde R'\left(z_1\rightarrow \frac{1}{z_1}, z_2\rightarrow \frac{z_2}{z_1}, z_3\rightarrow \frac{z_3}{z_1}\right)$.

\begin{align}
S_{10}=& -m_3^2\left(\frac{m_3^2}{4\pi\mu^2}\right)^{2(\nu-1)}\nonumber\\ 
\times &\left\{-\frac{z_1^\nu}{z_3} \Gamma^2(-\nu)\tilde M_M\left(1-\nu,1,1+\nu,1+\nu;\frac{z_1}{z_3},\frac{1}{z_3},\frac{z_2}{z_3}\right)\right. \nonumber\\
-&\frac{z_2^\nu}{z_3}\Gamma^2(-\nu)\tilde M_M\left(1,1-\nu,1-\nu,1+\nu;\frac{z_1}{z_3},\frac{1}{z_3},\frac{z_2}{z_3}\right)\nonumber\\
-&\frac{(z_1 z_2)^\nu}{z_3}  \Gamma^2(-\nu)\tilde M_M\left(1,1-\nu,1+\nu,1-\nu;\frac{z_1}{z_3},\frac{1}{z_3},\frac{z_2}{z_3}\right)\nonumber\\
+&\frac{z_1^\nu}{z_3}(-z_3)^\nu \frac{\Gamma^2(-\nu)\Gamma(\nu)\Gamma(1+\nu)}{\Gamma(2\nu)}\tilde M_M\left(1-2\nu,1-\nu,1+\nu,1-\nu;\frac{z_1}{z_3},\frac{1}{z_3},\frac{z_2}{z_3}\right)\nonumber\\
+&\frac{z_2^\nu}{z_3}(-z_3)^\nu \frac{\Gamma^2(-\nu)\Gamma(\nu)\Gamma(1+\nu)}{\Gamma(2\nu)}\tilde M_M\left(1-\nu,1-2\nu,1-\nu,1-\nu;\frac{z_1}{z_3},\frac{1}{z_3},\frac{z_2}{z_3}\right)\nonumber
\end{align}
\begin{align}
-&(-z_3)^{\nu-1}\frac{\Gamma^2(-\nu)\Gamma(\nu)\Gamma(1+\nu)}{\Gamma(2\nu)}\tilde M_M\left(1-2\nu,1-\nu,1-\nu,1+\nu;\frac{z_1}{z_3},\frac{1}{z_3},\frac{z_2}{z_3}\right)\nonumber\\
+&\left.(-z_3)^{2\nu-1}\frac{\Gamma^3(\nu)\Gamma(1-2\nu)}{\Gamma(3\nu)}\tilde M_M\left(1-3\nu,1-2\nu,1-\nu,1-\nu;\frac{z_1}{z_3},\frac{1}{z_3},\frac{z_2}{z_3}\right)\right\}
\label{S10}
\end{align}
is valid in the region $\tilde R'\left(z_1\rightarrow \frac{z_1}{z_2}, z_2\rightarrow \frac{1}{z_2}, z_3\rightarrow \frac{z_3}{z_2}\right)$.

\end{document}